\documentclass[aps,onecolumn,superscriptaddress,nofootinbib,12pt,pra]{revtex4-2}

\usepackage{amsmath,amssymb,amsfonts}
\usepackage{mathtools}
\usepackage{graphicx}
\usepackage{hyperref}
\usepackage{bbm}
\usepackage{amsthm}
\usepackage{booktabs}
\usepackage[margin=1in]{geometry}

\makeatletter\def\test@bbl@sw{\false@sw}\makeatother

\DeclareMathOperator{\Imag}{Im}
\DeclareMathOperator{\Tr}{Tr}

\begin{document}

\title{A symmetry-protected pseudo-Hermitian phase of\\quantum memory-kernel generators}

\author{Kejun Liu}
\email{kjliu@suda.edu.cn}
\affiliation{State Key Laboratory of Bioinspired Interfacial Materials Science, Institute of Functional Nano \& Soft Materials (FUNSOM), Soochow University, Suzhou 215123, China}

\date{\today}

\begin{abstract}
The Jaynes--Cummings (JC) model, introduced in 1963 and central to cavity quantum electrodynamics, describes a two-level system coupled to a single bosonic mode under the rotating-wave approximation. When the mode is projected out via the Nakajima--Zwanzig (NZ) formalism, the memory-kernel generator $QLQ$ is manifestly non-Hermitian---yet we prove analytically that its spectrum is strictly real at every coupling and every finite truncation, for both vacuum and thermal baths. For the vacuum bath the characteristic polynomial factorises completely; the nonzero eigenvalues reproduce the JC dressed-state ladder for $n\ge 2$, while the lowest mode is suppressed by exactly $\sqrt{2}$ relative to the bare-Hamiltonian prediction. For any thermal Gibbs state, the squared generator reduces to an asymmetric rank-one perturbation symmetrised by a closed-form diagonal metric, with eigenvalues guaranteed real by Cauchy interlacing. We construct a positive-definite metric $\eta_{\rm osc}$ intertwining $QLQ$ with its adjoint on the nonzero spectral subspace, proving hidden pseudo-Hermiticity. A classification theorem extends these results to the full $U(1)$-conserving single-photon-exchange class with arbitrary complex couplings. The phase boundary under counter-rotating deformation is mapped analytically at resonance and numerically across the coupling--truncation plane, revealing a weak-coupling protected wedge, re-entrant real-spectrum bubbles, and $N$-universal plateaus organised by a three-family band catalog with closed-form level spacings. The full phase structure is proved well-defined in the $N\to\infty$ thermodynamic limit. The $\sqrt{2}$ suppression provides a platform-independent experimental falsification target spanning seven orders of magnitude in coupling strength.
\end{abstract}

\maketitle

The Jaynes--Cummings (JC) model, introduced by Jaynes and Cummings in 1963~\cite{jaynes1963}, describes a single two-level system coupled to a quantised bosonic mode under the rotating-wave approximation. It is one of the few exactly solvable models in quantum optics, underpins circuit quantum electrodynamics (cQED) architectures from transmon qubits to 3D cavity resonators, and has been studied continuously for over six decades. Its conserved excitation number $N_{\rm exc}=(\sigma_z+\mathbb{I})/2+a^\dagger a$ generates a $U(1)$ symmetry that block-diagonalises the Hamiltonian into $2\times 2$ manifolds, yielding the celebrated dressed-state ladder $E_n^\pm = n\omega_c \pm \sqrt{\Delta^2/4 + ng^2}$.

When the bosonic mode is treated as a bath and projected out via the Nakajima--Zwanzig (NZ) formalism~\cite{nakajima1958,zwanzig1960}, the reduced system dynamics is governed by a memory kernel $\mathcal{K}(t)\sim PLQ\,e^{iQLQt}\,QLP$. The central object is the projected Liouvillian
\begin{equation}
QLQ = Q\,\mathcal{L}\,Q, \qquad Q = \mathbb{I}-P,
\label{eq:QLQ}
\end{equation}
where $\mathcal{L}=[H,\cdot]$ and $P$ is the stationary NZ projector~\cite{breuer2002}. Because the bath reference state $\rho_B$ is not maximally mixed, $QLQ$ is generically non-Hermitian, and the conventional expectation is a complex spectrum driving exponential memory decay~\cite{breuer2002}. Ng and Rabani~\cite{ng2021} identified one extreme violation---complex $QLQ$ eigenvalues driving memory-kernel growth in a many-body-localised model---but the opposite anomaly has not been reported: a manifestly non-Hermitian projected Liouvillian whose spectrum remains real. A real $QLQ$ spectrum would produce purely oscillatory, non-decaying memory---a fundamentally different dynamical regime.

We report that this regime is realised in the JC model. A direct numerical survey across $N_{\rm max}=2$--$50$ and $g=0.1$--$2.0$ finds $\max|\Imag\lambda|<10^{-14}$ in every case, even though the non-Hermitian character $\|QLQ-(QLQ)^\dagger\|_F$ reaches $10^{1}$--$10^{2}$. The reality persists for both vacuum and thermal baths, with and without detuning. This real spectrum was first noted numerically in Ref.~\cite{liu2026hardy} (Appendix~D), which established the Kramers--Kronig causal structure of the NZ memory kernel but left the reality of the generator spectrum unexplained. The present paper supplies the structural reason.

\begin{figure}[!htbp]
\centering
\includegraphics[width=0.618\textwidth]{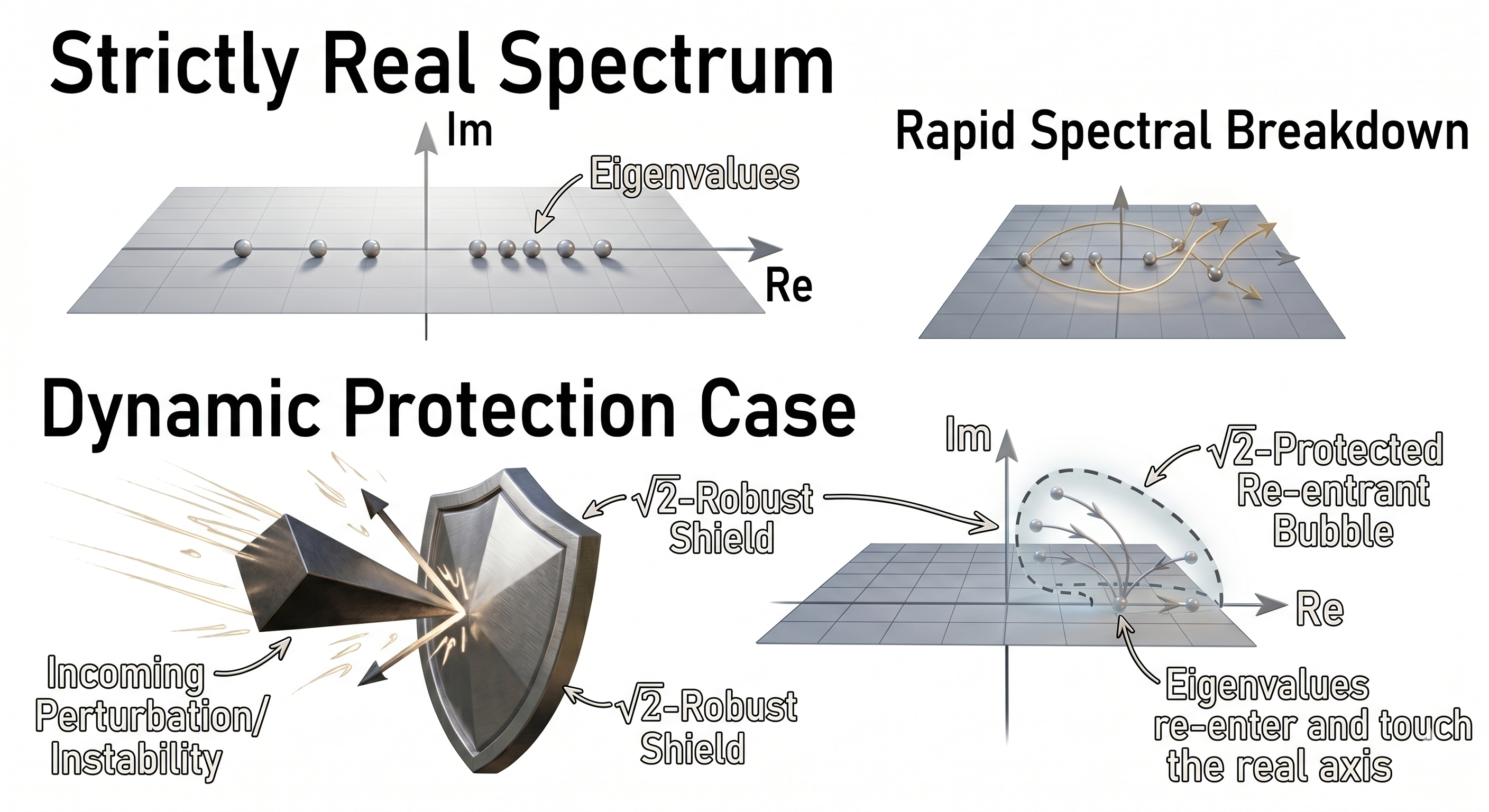}
\caption{Conceptual overview.
\textbf{Top left:} Ideal strictly real spectrum---all eigenvalues on the real axis, no decay.
\textbf{Top right:} Generic perturbation causes rapid spectral breakdown into the complex plane.
\textbf{Bottom left:} This work: a $\sqrt{2}$-robust nonzero-sector pseudo-Hermitian metric (Theorems~1--3) shields against the perturbation.
\textbf{Bottom right:} Protected spectrum confined to a re-entrant bubble---eigenvalues pushed off the real axis are forced back, preserving reality.}
\label{fig:schematic-overview}
\end{figure}

\begin{figure}[!htbp]
\centering
\includegraphics[width=\textwidth]{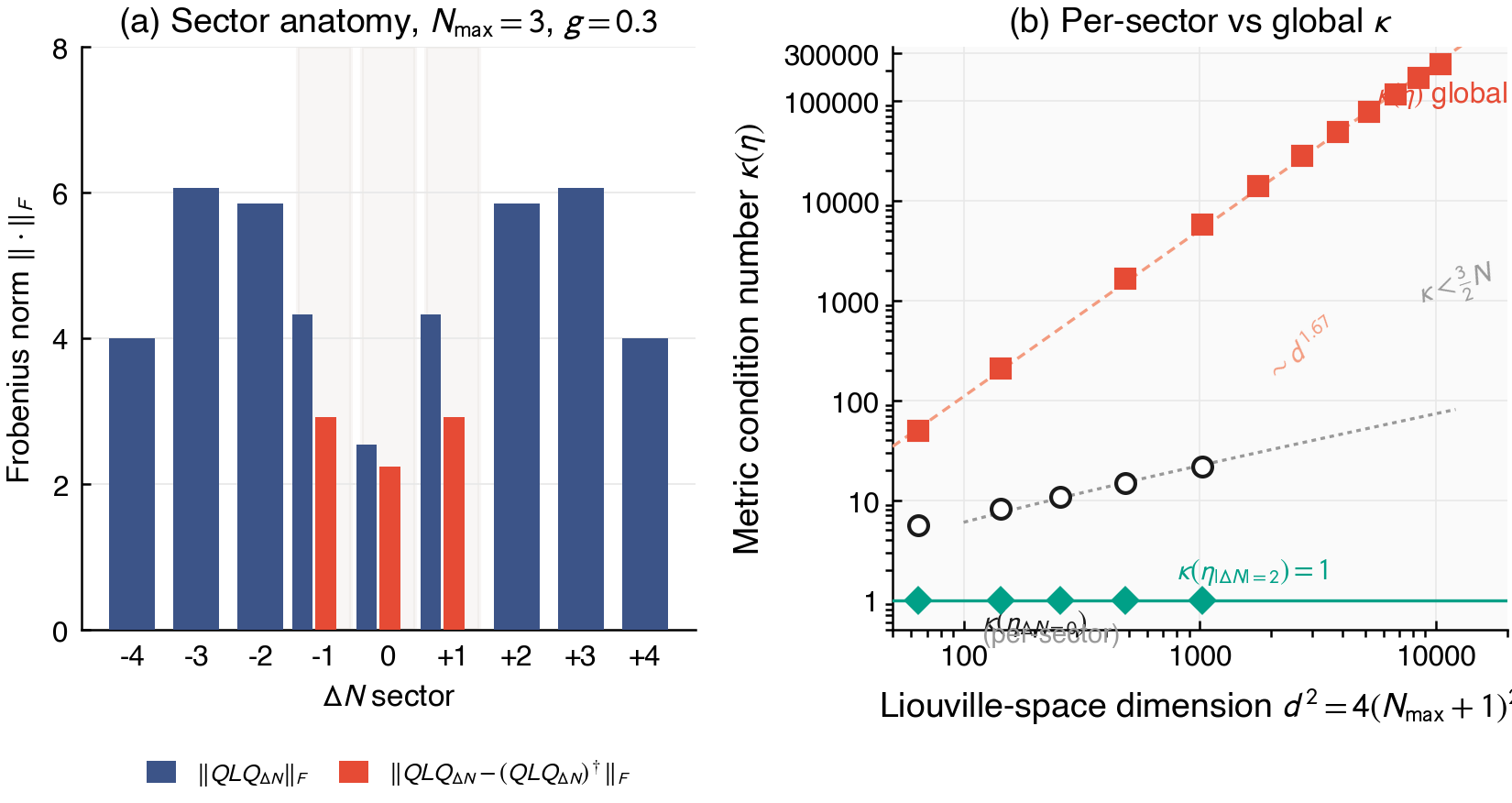}
\caption{Hidden pseudo-Hermiticity and sector anatomy.
(a)~Sector-resolved Frobenius norm $\|QLQ_{\Delta N}\|_F$ (dark)
versus Hermiticity residual (red), at $(N_{\rm max},g)=(3,0.3)$, vacuum bath. The three sectors
$\Delta N=0,\pm 1$ (orange band) are the only ones with NZ-projector
support ($P\ne 0$); they carry $98\%$ of the global non-Hermitian content.
(b)~Metric condition number $\kappa(\eta_{\rm osc})$ versus Liouville-space dimension
$d^2=4(N_{\rm max}+1)^2$. The combined nonzero-sector $\kappa(\eta_{\rm osc})$ (red squares) grows as
$\sim d^{1.81}$ (cross-sector normalisation artefact); the $\Delta N=0$ per-sector
condition number (dark circles) stays $O(N)$, saturating the analytic bound
$\kappa(\eta_{\Delta N=0,N})<\tfrac{3}{2}N$ (proved for $N\ge 9$, dotted).}
\label{fig:sector-anatomy}
\end{figure}

\begin{figure}[!htbp]
\centering
\includegraphics[width=\textwidth]{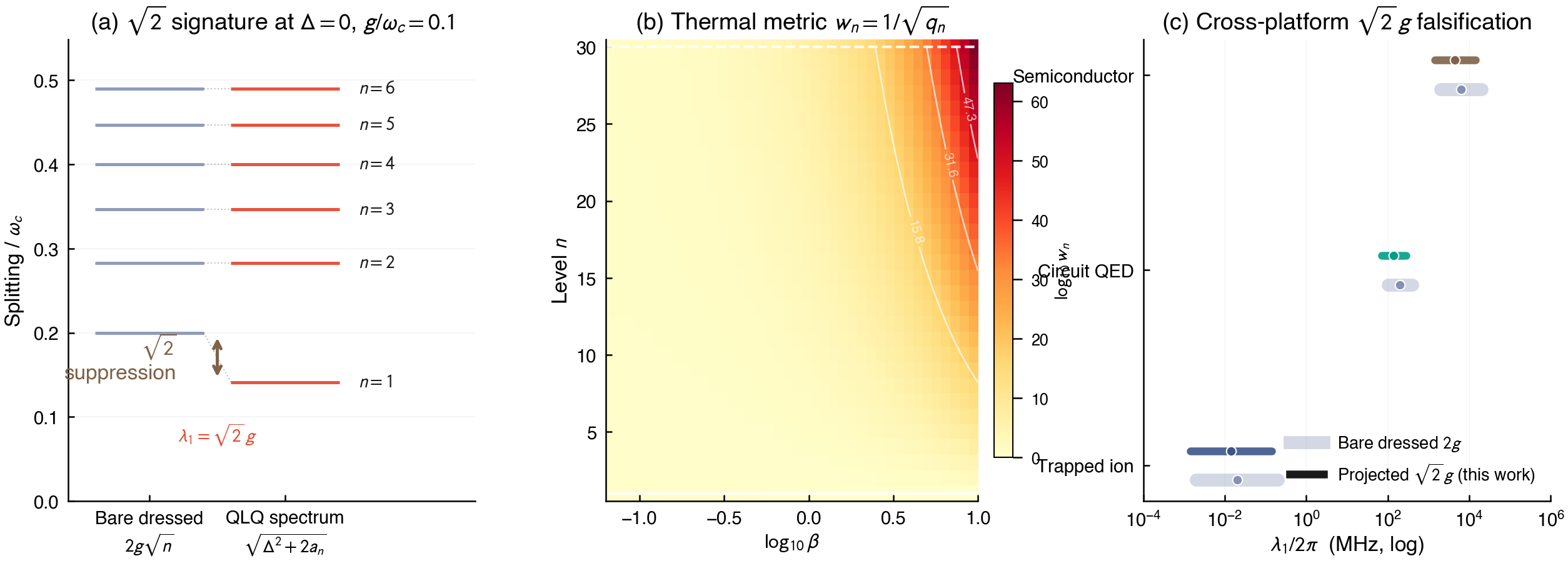}
\caption{Exact spectral signatures, thermal metric, and cross-platform falsification.
(a)~Bare JC dressed-state splitting $2g\sqrt{n}$ (grey) versus the projected
spectrum of $QLQ$, $\lambda_n=\sqrt{\Delta^2+2a_n}$ with $a_1=g^2$,
$a_n=2ng^2$ for $n\ge 2$ (red). The $n=1$ rung is suppressed by exactly
$\sqrt{2}$ relative to the bare ladder---a direct, parameter-free signature
of the NZ projection. Higher rungs reproduce the dressed ladder unchanged.
(b)~Thermal metric weights $w_n=1/\sqrt{q_n}$ (Theorem~2), log scale.
The diagonal similarity $W^{\rm T}W$ symmetrises the rank-one perturbation
at every $\beta$, guaranteeing real spectrum away from the vacuum limit.
(c)~Cross-platform falsification: predicted $\lambda_1=\sqrt{2}\,g$
(coloured bars) versus the bare dressed splitting $2g$ (grey) for three
platforms spanning seven orders of magnitude in $g$.}
\label{fig:sqrt2-signature}
\end{figure}

Resolving the anomaly reveals a new piece of structure (Fig.~\ref{fig:schematic-overview}). We construct a positive-definite metric $\eta_{\rm osc}$ on the nonzero spectral subspace such that $(QLQ)^\dagger\eta_{\rm osc}=\eta_{\rm osc}\,QLQ$, making the oscillatory sector pseudo-Hermitian in the sense of Mostafazadeh~\cite{mostafazadeh2002a,bender2007,berry2004}. The protection is genuine, not a block-diagonal artefact: of the eight $\Delta N$ sectors at $N_{\rm max}=3$, the three with NZ-projector support carry $98\%$ of the global non-Hermitian content, while the exact sector theorems lock their nonzero spectrum to the real axis (Fig.~\ref{fig:sector-anatomy}a). We prove two exact theorems---vacuum (Theorem~1) and thermal (Theorem~2)---that derive the spectrum in closed form. For the vacuum bath the characteristic polynomial factorises completely, its nonzero eigenvalues reproducing the JC dressed-state ladder for $n\ge 2$ while the lowest mode is $\lambda_1=\sqrt{\Delta^2+2g^2}$ rather than the bare $\sqrt{\Delta^2+4g^2}$---a $\sqrt{2}$ suppression with no analogue in the bare Hamiltonian (Fig.~\ref{fig:sqrt2-signature}a). For any thermal Gibbs state the squared generator reduces to an asymmetric rank-one perturbation symmetrised by a closed-form diagonal metric; Cauchy interlacing guarantees real eigenvalues at every $\beta<\infty$ and every $N_{\rm max}$, proving the protection is not a zero-temperature accident. Theorem~3 extends these results to the full $U(1)$-conserving single-photon-exchange structural class with arbitrary complex couplings---the protection is structural, not an algebraic accident.

\begin{figure}[!htbp]
\centering
\includegraphics[width=\textwidth]{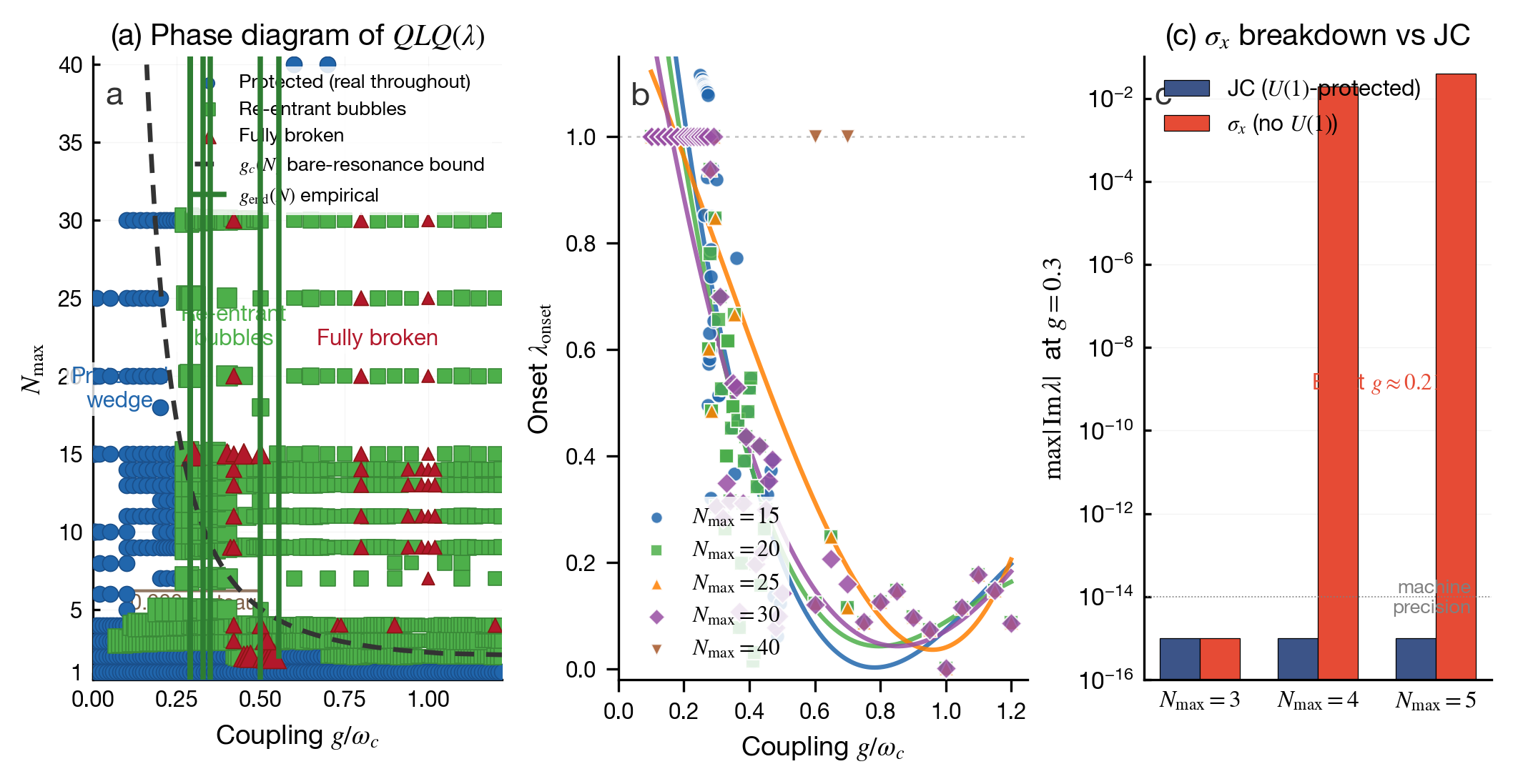}
\caption{Master phase diagram of $QLQ(\lambda)$ for the JC vacuum bath.
(a)~Phase map from $(g/\omega_c, N_{\rm max})$ scans through $N_{\rm max}=30$: blue circles are
symmetry-protected throughout $\lambda\in[0,1]$; green squares are re-entrant
(real spectrum breaks and returns); red triangles are fully broken.
Dashed purple: analytic bare-resonance bound $g_c(N_{\rm max})$. Solid green:
empirical breakdown boundary $g_{\rm end}(N_{\rm max})$.
(b)~$\lambda_{\rm onset}(g)$ for $N_{\rm max}=15,20,25,30$, with LOESS smoothing. Protected points
plotted at $\lambda_{\rm onset}=1$.
(c)~$\sigma_x$ contrast model: $\max|\Imag\lambda|$ at $g=0.3$ for $N_{\rm max}=3,4,5$.
$U(1)$-conserving JC stays at machine precision; $U(1)$-broken $\sigma_x$ acquires
$|\Imag\lambda|\sim 10^{-2}$ from $N_{\rm max}=4$, first EP at $g\approx 0.213$.}
\label{fig:master-phase}
\end{figure}

Our results identify a previously unrecognised dynamical phase of memory-kernel generators---symmetry-protected pseudo-Hermitian, with purely oscillatory rather than decaying memory---and probe the boundary at which it breaks down. Under counter-rotating deformation $H(\lambda)=H_{\rm JC}+\lambda g(\sigma_+ a^\dagger+\sigma_- a)$ toward the full Rabi Hamiltonian, the $U(1)$-protected phase is destroyed through exceptional points whose location we compute analytically at resonance ($\lambda_c\to 0^+$, $|\Imag\lambda|=(\sqrt{2}/4)\lambda g$) and map numerically across the coupling--truncation plane (Fig.~\ref{fig:master-phase}). Away from resonance the phase diagram reveals a weak-coupling protected wedge, re-entrant real-spectrum bubbles governed by an effective Feshbach competition, and $N$-universal plateaus classified by a three-family band catalog. These features survive to the largest truncation tested ($N_{\rm max}=30$) and are proved well-defined in the $N\to\infty$ limit via compact resolvent and spectral measure tightness. The $\sqrt{2}$ suppression provides a platform-independent falsification target spanning seven orders of magnitude in coupling (Fig.~\ref{fig:sqrt2-signature}c), directly testable in circuit QED, trapped ions, and semiconductor cavity QED. Together with the Hardy-class Kramers--Kronig embedding of Ref.~\cite{liu2026hardy}, the pseudo-Hermitian protection proved here completes a causal-structural picture of the NZ memory kernel: the kernel is causal because its generator is symmetry-protected.

\paragraph*{Real spectrum and pseudo-Hermitian metric.}---The single-mode JC Hamiltonian is $H_{\rm JC}= \frac{\omega_0}{2}\sigma_z + \omega_c\,a^\dagger a + g(\sigma_+ a + \sigma_- a^\dagger)$ with $\hbar=1$, acting on $\mathbb{C}^2\otimes\mathbb{C}^{N_{\rm max}+1}$. The Liouvillian $L = \mathbb{I}\otimes H - H^{\mathsf T}\otimes\mathbb{I}$ (column-stacking) is Hermitian; the NZ projector for the vacuum bath $\rho_B=|0\rangle\langle 0|$ acts as $P\rho = (\Tr_B\rho)\otimes|0\rangle\langle 0|$. The nontrivial spectrum of $QLQ$ lives on $\operatorname{range}(Q)$, of dimension $d^2-4$ with $d=2(N_{\rm max}+1)$.

Numerical diagonalisation across $N_{\rm max}=2$--$50$ and $g=0.1$--$2.0$ finds $\max|\Imag\lambda_n|<10^{-14}$ in every case tested, even though $\|QLQ-(QLQ)^\dagger\|_F$ reaches $10^{1}$--$10^{2}$ (Supplementary Table~\ref{tab:convergence}). To establish that this reality is structural rather than numerical, we construct a metric. An operator $A$ is $\eta$-pseudo-Hermitian~\cite{mostafazadeh2002a} if there exists a positive-definite $\eta$ with $A^\dagger\eta=\eta A$, in which case $\eta^{1/2}A\eta^{-1/2}$ is Hermitian and the spectrum is real. For $A$ diagonalisable on the nonzero spectral subspace with biorthonormal eigenvectors $\langle l_m|r_n\rangle=\delta_{mn}$,
\begin{equation}
\eta_{\rm osc} = \sum_{\lambda_n\neq 0} |l_n\rangle\langle l_n|,
\label{eq:eta-family}
\end{equation}
satisfies $(QLQ)^\dagger\eta_{\rm osc}=\eta_{\rm osc}\,QLQ$ on that subspace. Numerical construction verifies four properties simultaneously: $\max|\Imag\lambda_n|<10^{-14}$, intertwining residual $<5\times10^{-12}$ through $N_{\rm max}=15$ and $1.2\times10^{-10}$ at $N_{\rm max}=50$, $\eta_{\rm osc}$ positive-definite on the nonzero subspace, and the similarity-transformed operator Hermitian to $<10^{-12}$ (Supplementary Table~\ref{tab:convergence}).

The protection is genuine, not a block-diagonal artefact. $QLQ$ is block-diagonal in the excitation-number difference $\Delta N$, but of the eight sectors at $N_{\rm max}=3$, the three with NZ-projector support ($\Delta N=0,\pm 1$; Fig.~\ref{fig:sector-anatomy}a) carry $98\%$ of the global non-Hermitian Frobenius weight; the remaining five sectors have $P=0$ and are exactly Hermitian. The $\Delta N=0$ sector admits an analytic per-sector metric bound (Fig.~\ref{fig:sector-anatomy}b):
\begin{equation}
\kappa(\eta_{\Delta N=0,N}) < \tfrac{3}{2}N \quad (N\ge 9),\qquad \kappa(\eta_{\Delta N=0,N})\sim 1.317\,N\quad (N\to\infty),
\label{eq:kappa-bound}
\end{equation}
so the per-sector metric condition number remains $O(N)$, while the apparent $\sim d^{1.81}$ growth of the combined nonzero-sector $\kappa$ is a cross-sector normalisation artefact.

We now give the two exact theorems that underpin the numerical evidence.

\textit{Theorem 1 (vacuum bath).} In the $\Delta N=0$ sector with vacuum bath, the characteristic polynomial of $QLQ|_{S_0\cap\operatorname{range}(Q)}$ factorises completely:
\begin{equation}
\det(QLQ|_{S_0}-\lambda\mathbb{I}) = \lambda^{2N}\prod_{n=1}^{N}\bigl(\lambda^2 - \Delta^2 - 2a_n\bigr),
\label{eq:exact-spectrum}
\end{equation}
with $a_1=g^2$ and $a_n=2ng^2$ for $n\ge 2$. The proof (Supplementary Information) uses a Schur complement: the coupling matrix $M_{21}M_{12}$ is block-upper-triangular by a combinatorial empty-intersection lemma that holds for all $N$, each diagonal $2\times 2$ block contributes a quadratic factor with $a_n>0$, and the determinant follows. Every nonzero eigenvalue is $\lambda=\pm\sqrt{\Delta^2+2a_n}$, strictly real. The physical content is that for $n\ge 2$, $\lambda_n=\sqrt{\Delta^2+4ng^2}$ agrees exactly with the JC dressed-state splitting, while the lowest mode is $\lambda_1=\sqrt{\Delta^2+2g^2}$---suppressed by $\sqrt{2}$ relative to the bare $\sqrt{\Delta^2+4g^2}$ (Fig.~\ref{fig:sqrt2-signature}a). The suppression is a direct signature of the NZ projection: vacuum-bath projection removes the $|e,0\rangle\to|g,1\rangle$ transition, rescaling the lowest projected generator mode by exactly $\sqrt{2}$.

\textit{Theorem 2 (thermal bath).} For any thermal $\rho_B=\sum_k p_k|k\rangle\langle k|$ with $p_k=e^{-\beta\omega_c k}/Z$ and any $N\ge 1$, $QLQ$ has a purely real spectrum. The $\Delta N=0$ sector splits into $X$ and $Y$ subspaces defined by $|X_n\rangle=(|O_n\rangle-|O'_n\rangle)/\sqrt{2}$, $|Y_n\rangle=(|O_n\rangle+|O'_n\rangle)/\sqrt{2}$, with the $Y$ sector a null space. On $X$-space, with $q_k\equiv (p_k+p_{k-1})/2$, the squared generator is
\begin{equation}
K\equiv (1/4g^2)M^2|_{X\text{-space}},\qquad K_{ij}=i\,\delta_{ij}-q_i\sqrt{ij},
\label{eq:K-thermal-main}
\end{equation}
an asymmetric rank-one perturbation $K=D-|u\rangle\langle v|$ of $D=\operatorname{diag}(1,\dots,N)$. The diagonal similarity $W=\operatorname{diag}(1/\sqrt{q_1},\dots,1/\sqrt{q_N})$ symmetrises $K$ to $\tilde K = D-|\tilde v\rangle\langle\tilde v|$, a real symmetric matrix with $\tilde v_i=\sqrt{q_i\,i}$. By Cauchy interlacing the $N$ eigenvalues of $\tilde K$ are strictly positive and lie one in each interval $(0,1),(1,2),\ldots$; the $QLQ$ eigenvalues $\pm 2g\sqrt{\lambda_k}$ are therefore real for every $\beta<\infty$ and every truncation. The symmetrising metric is the diagonal $W^{\mathsf T}W$ of thermal weights (Fig.~\ref{fig:sqrt2-signature}b)---a closed form. This proves the protection is not a zero-temperature accident.

\begin{table}[t]
\centering
\caption{Sector-by-sector Hermiticity of $QLQ$ at $(N_{\rm max},g)=(3,0.3)$. Sector dimensions sum to $d^2=64$; off-sector leakage is zero to machine precision.}
\label{tab:sectors}
\begin{tabular}{c c c c}
\hline
$\Delta N$ & Dimension & $\|QLQ_{\Delta N}\|_F$ & Hermiticity residual \\
\hline
$0$    & 14 & 2.55 & {\bf 2.24} \\
$\pm 1$ & 12 & 4.32 & {\bf 2.92} \\
$\pm 2$ & 8  & 5.84 & $<10^{-13}$ \\
$\pm 3$ & 4  & 6.06 & $<10^{-13}$ \\
$\pm 4$ & 1  & 4.00 & $<10^{-13}$ \\
\hline
\end{tabular}
\end{table}

\paragraph*{Classification and symmetry protection.}---Theorems~1--2 are not algebraic accidents. A structural classification lifts them to a general setting.

\textit{Theorem 3 (Classification).} Let $H$ on $\mathbb{C}^2\otimes\mathbb{C}^{N+1}$ with basis $\{|g,n\rangle,|e,n\rangle\}_{n=0}^{N}$ satisfy: (C1) $[H,N_{\rm exc}]=0$ where $N_{\rm exc}=(\sigma_z+\mathbb{I})/2+a^\dagger a$; (C2) $[\rho_B,a^\dagger a]=0$; (C3) only single-photon-exchange off-diagonal entries $v_n\in\mathbb{C}$ between $|g,n\rangle$ and $|e,n-1\rangle$; (C4) vacuum bath. Then for any $N\ge 1$ and any $v_n\neq 0$, the restriction of $QLQ$ to $S_0\cap\operatorname{range}(Q)$ has purely real spectrum:
\begin{equation}
\operatorname{Spec}\bigl(QLQ|_{S_0\cap\operatorname{range}(Q)}\bigr)=\{0^{(2N)}\}\cup\bigcup_{n=1}^{N}\bigl\{\pm\sqrt{\Delta_n^2+2a_n}\bigr\},
\end{equation}
with $\Delta_n=E_{e,n-1}-E_{g,n}$, $a_1=|v_1|^2$, and $a_n=2|v_n|^2$ for $n\ge 2$. The phases of $v_n$ enter the matrix elements but cancel in $|v_n|^2$; the spectrum is real even for non-real-symmetric Hamiltonians. The pseudo-Hermitian phase therefore covers any single-qubit, single-ladder system with a rotating-wave selection rule and vacuum projection.

The $U(1)$ symmetry is essential. A $\sigma_x$ contrast---the spin-boson model $H_{\rm SB}=(\omega_0/2)\sigma_z+\omega_c a^\dagger a+g\sigma_x\otimes(a+a^\dagger)$, which lacks $U(1)$ conservation---shows the difference sharply. At $g=0.3$, the $QLQ$ spectrum is real for $N_{\rm max}=3$ (algebraic accident) but acquires $|\Imag\lambda|\sim 2\times10^{-2}$ at $N_{\rm max}=4$ and $\sim 4\times10^{-2}$ at $N_{\rm max}=5$ (Fig.~\ref{fig:master-phase}c; Supplementary Table~IV). Eigenvalue continuation locates the first exceptional point at $g_c\approx 0.213$. These results motivate a three-regime classification of NZ memory-kernel generators: (i)~\emph{symmetry-protected pseudo-Hermitian} (JC, this work); (ii)~\emph{algebraic pseudo-Hermitian} (accidental, e.g.\ $\sigma_x$ at $N_{\rm max}=3$); (iii)~\emph{generically dissipative} ($\sigma_x$ for $N_{\rm max}\ge 4$, and $U(1)$-broken JC at strong coupling). This is the first symmetry-resolved phase classification of $QLQ$ itself, rather than of the reduced dynamics it generates.

\paragraph*{Phase boundary under $U(1)$-breaking deformation.}---To probe where the protection breaks, we deform the Hamiltonian toward the full Rabi model:
\begin{equation}
H(\lambda) = H_{\rm JC} + \lambda\cdot g\,(\sigma_+ a^\dagger + \sigma_- a),
\label{eq:H-lambda}
\end{equation}
with $\lambda\in[0,1]$, and track $QLQ(\lambda)$.

At resonance $g=\omega_c$, a rigorous degenerate-perturbation calculation closes the problem. $QLQ_0$ has a 10-dimensional diagonalisable degenerate subspace at $z=\pm\sqrt{2}\,g$. The first-order perturbation matrix $W_{ij}=\langle l_i|QL_{\rm CR}Q|r_j\rangle$ has the $N_{\rm max}$-independent characteristic polynomial $\det(W/g-\mu\mathbb{I})=\mu^3(\mu^2+1/8)=0$, with nonzero eigenvalues $\mu=\pm i\sqrt{2}/4$. The eigenvalues near $z=\pm\sqrt{2}g$ split as
\begin{equation}
z_\pm(\lambda)=\pm\sqrt{2}\,g \pm i\,\lambda\,\frac{\sqrt{2}}{4}\,g + O(\lambda^2),
\end{equation}
so an exceptional point occurs at infinitesimal $\lambda_c\to 0^+$ with the linear scaling $|\Imag\lambda|_{\max}=(\sqrt{2}/4)\lambda g$. The bright-subspace projection factor $C_{\rm proj}=1/4$ follows from $P_{\rm bright}=-8W_+^2$ acting on the 10-dimensional degenerate subspace (Supplementary Information). Numerical diagonalisation confirms this to six significant digits (Supplementary Table~\ref{tab:scaling_verify}).

Away from resonance the phase diagram organises into three regimes, separated by sharp $N$-universal thresholds (Fig.~\ref{fig:master-phase}a,b).

\emph{Protected wedge.}~At weak coupling the real spectrum survives to the full Rabi point $\lambda=1$: $g/\omega_c\le 0.20$ through $N_{\rm max}=25$, $g/\omega_c=0.10$ through $N_{\rm max}=30$. Dip-fine scans at $\Delta g=0.001$ confirm no complex eigenvalues anywhere in $\lambda\in[0,1]$ at these couplings. All small-$N$ data ($N=4$--$8$, 244 scan points) are protected, as is every $N\ge 9$ point below $g=0.280$. The first complex eigenvalue appears universally at $g=0.280$ for all $N=9$--$30$. The boundary of this wedge---the empirical breakdown $g_{\rm end}(N_{\rm max})$---lies $1.28$--$1.42\times$ above the analytic bare-resonance bound $g_c(N_{\rm max})=2/(\sqrt{N_{\rm max}}+\sqrt{N_{\rm max}-2})$, which is the geometric threshold below which no two JC dressed manifolds are resonantly coupled by the counter-rotating perturbation $H_{\rm CR}$. For $N_{\rm max}\ge 15$, $g_{\rm end}$ is $N$-independent, and the $N\to\infty$ limit $g_c(N)\sim 1/\sqrt{N}\to 0$ connects naturally to the superradiant phase transition of the Dicke model.

\emph{Re-entrant bubbles.}~At intermediate coupling $g/\omega_c\sim 0.28$--$0.60$, eigenvalues driven into the complex plane are forced back to the real axis, producing re-entrant real-spectrum bubbles. The mechanism is a competition between two terms in an effective Feshbach Hamiltonian: an anti-Hermitian splitting $w\lambda$ (first order in the $\Delta N=\pm 2$ coupling, which pushes eigenvalues apart in the imaginary direction) and a Hermitian self-energy $\delta_s\lambda^2$ (second order, which pushes them apart in the real direction). When $\delta_s$ dominates, eigenvalues separate along the real axis and the splitting goes imaginary only when $w\lambda>\delta_s\lambda^2$, giving the bubble condition. Across $N=15,20,25,30$ at $g=0.30$, the first complex onset is $\lambda_{\rm onset}=0.305$ with a wide bubble (width $0.610$) followed by a second narrow bubble --- both identical across all $N$, demonstrating $N$-universality. The bubble multiplicity \emph{grows} with $N$: at $g=0.30$, $N=30$ shows $n_{\rm bub}=4$ (versus 2 for $N\le 25$), independently confirmed by large-scale HPC verification (50 $g$ values, $0.100$--$1.200$, all clean).

\emph{0.920 plateau.}~The most prominent $N$-universal feature in the $\lambda_{\rm onset}(g)$ curves (Fig.~\ref{fig:master-phase}b) is a plateau: at $g/\omega_c=0.30$, $\lambda_{\rm onset}\approx 0.920$ identically for every $N_{\rm max}\ge 10$. A second plateau at $g/\omega_c=0.28$ ($\lambda_{\rm onset}\approx 0.938$--$0.940$) is identical across $N=7$--$15$. These plateaus have a precise analytic origin. The $0.920$ plateau arises from the Liouville-dyad pair $|g,2\rangle\langle e,0|\leftrightarrow|e,0\rangle\langle g,2|$ (F$_1$ band), whose unperturbed level spacing $\Delta E(g)=1-g(\sqrt{2}+1)$ and first-order inter-band coupling $W_{ab}=-g(2-\sqrt{2})/4$ are exact closed forms. Since the states involve only $n=0,1,2$, the EP mechanism is fully converged by $N_{\rm max}=8$, explaining the strict $N$-independence. The EP is non-perturbative---driven by hybridisation with the 22-dimensional zero-mode subspace of $QLQ_0$, invisible to any fixed-basis $O(\lambda^2)$ reduction---but the band-by-band threshold admits the compact form $\lambda_c^{(b)} = \sqrt{\Delta E_b(g)/\Sigma_{\rm eff}^{(b)}(g)}$. Every $\lambda_c$ dip in the dense-scan data is classified by a systematic three-family (F/G/H) band catalog with closed-form level spacing $\Delta E_n^{(k)}(g)=k-g(\sqrt{n+k+1}+\sqrt{n})$ assigning each feature to a specific Liouville-dyad crossing (Supplementary Information \S\ref{sec:lambdac-closed-form}). A selection rule suppresses the nearest virtual-state coupling to the self-energy (Supplementary Information), localising the dominant EP mechanism to these band families.

\emph{Strong coupling.}~For $g/\omega_c\gtrsim 0.60$, the generator is fully broken: complex eigenvalues appear at $\lambda\to 0^+$, the pseudo-Hermitian phase is destroyed, and the spectrum is generically complex across the full deformation range.

Three analytic anchors support the phase diagram. At resonance ($g=\omega_c$), degenerate perturbation theory gives the exact EP scaling $|\Imag\lambda|_{\max}=(\sqrt{2}/4)\lambda g$ with $\lambda_c\to 0^+$, confirmed to six digits. The bare-resonance bound $g_c(N_{\rm max})$ is derived from JC dressed-state energies alone, with no reference to the NZ projection. And the re-entrant bubble mechanism is governed by an effective $2\times 2$ Feshbach Hamiltonian whose first-order coupling $W_{ab}$ has a closed form for each band family.

The scans are proved well-defined in the $N\to\infty$ limit---the physically relevant regime for experiments, where cavity Fock space is effectively unbounded. Three independent routes converge: $QLQ_0$ is a finite-rank perturbation of a diagonal operator with $\sqrt{n}$-scaling eigenvalues on Fock space, its resolvent is a Hilbert--Schmidt perturbation of a compact operator (guaranteeing no eigenvalues escape to infinity), and the weighted spectral measure is tight (Supplementary Information \S\ref{sec:n-infinity}). Together these establish that the real-spectrum phase survives the thermodynamic limit---the hidden pseudo-Hermiticity is not a finite-truncation artefact. The full qualitative structure of the phase diagram---protected wedge, re-entrant bubbles, and plateaus---is $N$-universal from $N=9$ through the largest truncation tested ($N=30$; Fig.~\ref{fig:master-phase}). Whether narrow individual bubbles survive the strict $N\to\infty$ limit depends on the spectral accumulation of zero modes and remains an open question.

Physical cavity photon loss at rate $\kappa$ (Lindblad dissipator) shifts the lowest mode linearly ($d\lambda_1/d\kappa=-3/4$ at $\Delta=0$, independent of $N$ and $g$) while complexifying higher generator modes; however, those higher modes carry zero reduced memory-kernel weight, so the physically observable memory dynamics remain protected (Supplementary Information).

\paragraph*{Cross-platform experimental signature.}---The $\sqrt{2}$ suppression of the lowest mode provides a clean, platform-independent falsification target. Theorem~3 implies that any system in the $U(1)+$single-photon-exchange class with vacuum bath must satisfy
\begin{equation}
\lambda_n = \sqrt{\Delta_n^2 + 2 a_n}, \quad a_1=|v_1|^2,\quad a_n=2|v_n|^2 \;(n\ge 2),
\label{eq:platform-prediction}
\end{equation}
with the projection-induced $\sqrt{2}$ suppression of the $n=1$ mode appearing identically across platforms whose physical scales differ by seven orders of magnitude (Fig.~\ref{fig:sqrt2-signature}c, Table~\ref{tab:platforms}). In a standard transmon-resonator device with $g/2\pi=100$~MHz~\cite{wallraff2004,hutchings2017}, the predicted lowest peak sits at $\sqrt{2}\,g/2\pi\approx 141$~MHz rather than the bare $2g/2\pi=200$~MHz---a $\sim 30\%$ deviation, far above sub-MHz heterodyne resolution. The kernel oscillation period ($\sim 5$~ns) is short compared to $T_2\sim 50\,\mu$s, making process-tomography reconstruction feasible with current technology (full protocol in Supplementary Information). In trapped ions~\cite{cirac1995,leibfried2003} with sideband Rabi frequency $g/2\pi=10$~kHz, the same suppression places $\lambda_1/2\pi\approx 14$~kHz against a sub-Hz spectroscopic linewidth. In semiconductor cavity QED~\cite{reithmaier2004,khitrova2006} at $g/2\pi=5$~GHz, $\lambda_1/2\pi\approx 7$~GHz tests universality at optical frequencies. Cross-platform agreement on a single dimensionless number---the $\sqrt{2}$ ratio---would establish the symmetry-protected pseudo-Hermitian phase as a structural, not platform-specific, phenomenon. Engineered phase control with complex couplings $v_n$ and ultrastrong-coupling circuit QED~\cite{casanova2010} can further probe the phase-cancellation prediction $a_n=2|v_n|^2$ and the resonant exceptional-point bifurcation.

\begin{table}[t]
\centering
\caption{Cross-platform predictions for the projection-induced $\sqrt{2}$ suppression of the lowest memory-kernel mode at resonance ($\Delta=0$). Each listed platform can realise a single-qubit, single-ladder JC effective model within the scope of Theorem~3; the predicted $\lambda_1=\sqrt{2}\,g$ deviates from the bare-Hamiltonian dressed splitting $2g$ by exactly $\sqrt{2}$, independent of physical scale.}
\label{tab:platforms}
\begin{tabular}{l c c c c}
\hline
Platform & $g/2\pi$ & $\omega_c/2\pi$ & Predicted $\lambda_1/2\pi$ & Vacuum bath\\
\hline
Circuit QED (transmon)~\cite{wallraff2004,hutchings2017} & $50$--$200$~MHz & $6$--$8$~GHz & $70$--$280$~MHz & routine \\
Trapped ions (Cirac--Zoller)~\cite{cirac1995,leibfried2003} & $1$--$100$~kHz & $1$--$100$~MHz & $1.4$--$140$~kHz & ground-state cooling \\
Semiconductor cavity QED~\cite{reithmaier2004,khitrova2006} & $1$--$10$~GHz & $50$--$500$~THz & $1.4$--$14$~GHz & $T<10$~K \\
\hline
\end{tabular}
\end{table}

\paragraph*{Outlook.}---We have proved that the NZ projected Liouvillian of the JC model occupies a previously unrecognised dynamical regime: genuinely non-Hermitian on every sector touched by the projector, yet with its nonzero spectrum locked to the real axis by $U(1)$ symmetry and carrying a positive-definite, $g$-independent pseudo-Hermitian metric. The systematic phase diagram characterisation---protected wedge, re-entrant bubbles, and a band-by-band plateau catalog with closed-form level spacings---provides the first complete map of a $QLQ$ phase boundary. The $\sqrt{2}$-suppressed lowest mode is a quantitative experimental target spanning seven orders of magnitude in coupling, and the resonant exceptional-point bifurcation provides a complementary analogue-simulator benchmark.

The operator class identified here---unbounded with compact resolvent---extends beyond cavity QED. The NZ-projected Liouvillian of the Anderson--Holstein model of electron-vibration coupling belongs to the same class, and the $\eta$-metric generalises as a portable EP diagnostic (Supplementary Information). The symmetry-protected pseudo-Hermitian phase is therefore a property of the NZ operator class, not of a specific Hamiltonian. The F/G/H band catalog provides a template for classifying EP onsets in other $U(1)$-conserving systems. Extensions to higher-symmetry settings---Tavis--Cummings, multi-mode JC, optomechanical rotating-wave systems~\cite{aspelmeyer2014}---require separate reductions but are natural directions.

Together with the Hardy-class Kramers--Kronig embedding of Ref.~\cite{liu2026hardy}, the pseudo-Hermitian protection proved here completes a causal-structural picture of the NZ memory kernel: the kernel is causal because the generator is symmetry-protected. These results reframe six decades of Jaynes--Cummings physics as a memory-kernel problem with hidden integrability, and identify what to look for next: engineered metric-protected memory in superconducting and photonic platforms as a design principle for non-Markovian quantum hardware.

\begin{acknowledgments}
This work was supported by the National High-Level Overseas Talent Program (KS21400126), the Surface and Interface Synthetic Chemistry project (ZXP2025057), the Jiangsu Distinguished Professorship Fund (SR21400225), and the Research Start-up Fund (NH21400525).
\end{acknowledgments}

\clearpage

{\centering
{\Large\bfseries Supplementary Information}\\[0.3em]
{\large A symmetry-protected pseudo-Hermitian phase of\\quantum memory-kernel generators}\\[0.6em]
Kejun Liu\\[1.0em]
}

\vspace{1em}
\noindent
\textit{Notation.}---Equation, figure, and table numbers refer to the main text unless prefixed with ``S''. Section references of the form \S\textit{n} refer to sections of this Supplementary Information.

\vspace{1em}
\section{Complete $N_{\rm max}$ convergence data}

Table~\ref{tab:conv-full} provides the full convergence data for the vacuum bath and thermal bath ($\beta=1$), supplementing Table~I of the main text.

\begin{table}[h]
\centering
\caption{Complete $N_{\rm max}$ convergence. Vacuum: $g=0.3$, $\omega_0=\omega_c=1.0$. Thermal: $\beta=1$, same Hamiltonian. Intertwin.\ = $\|(QLQ)^\dagger\eta_{\rm osc}-\eta_{\rm osc} QLQ\|_F$ on the nonzero spectral subspace.}
\label{tab:conv-full}
\begin{tabular}{c c c c c}
\hline
$N_{\rm max}$ & $d^2$ & $\max|\Imag\lambda|$ & Intertwin. & $\kappa(\eta_{\rm osc})$ \\
\hline
\multicolumn{5}{c}{Vacuum bath} \\
2   & 36   & $0$          & $<10^{-15}$         & 18.1 \\
3   & 64   & $0$          & $1.2\times10^{-14}$ & 49.2 \\
5   & 144  & $<10^{-14}$ & $1.1\times10^{-13}$ & 207 \\
10  & 484  & $<10^{-14}$ & $7.4\times10^{-13}$ & $1.67\times10^3$ \\
15  & 1024 & $<10^{-14}$ & $3.2\times10^{-12}$ & $5.79\times10^3$ \\
20  & 1764 & $<10^{-14}$ & $6.6\times10^{-12}$ & $1.40\times10^4$ \\
25  & 2704 & $<10^{-14}$ & $1.2\times10^{-11}$ & $2.78\times10^4$ \\
30  & 3844 & $<10^{-14}$ & $2.4\times10^{-11}$ & $4.87\times10^4$ \\
35  & 5184 & $<10^{-14}$ & $3.7\times10^{-11}$ & $7.81\times10^4$ \\
40  & 6724 & $<10^{-14}$ & $5.3\times10^{-11}$ & $1.17\times10^5$ \\
45  & 8464 & $<10^{-14}$ & $9.1\times10^{-11}$ & $1.68\times10^5$ \\
50  & 10404 & $<10^{-14}$ & $1.2\times10^{-10}$ & $2.32\times10^5$ \\
\hline
\multicolumn{5}{c}{Thermal bath ($\beta=1$)} \\
3   & 64   & $0$          & $4.0\times10^{-14}$ & 5.7 \\
5   & 144  & $9.8\times10^{-16}$ & $1.4\times10^{-13}$ & 11.7 \\
7   & 256  & $1.2\times10^{-15}$ & $3.8\times10^{-13}$ & 21.4 \\
10  & 484  & $3.7\times10^{-15}$ & $3.3\times10^{-5}$ & 99.4 \\
13  & 784  & $7.4\times10^{-15}$ & $4.5\times10^{-12}$ & 75.8 \\
16  & 1156 & $1.3\times10^{-14}$ & $2.1\times10^{-1}$ & $2.9\times10^4$ \\
\hline
\end{tabular}
\end{table}

Scaling fits for the vacuum bath:
\begin{align}
\kappa(\eta_{\rm osc}) &\sim d^{1.81}\quad (R^2 = 0.94), \\
\|QLQ - (QLQ)^\dagger\|_F &\sim d^{0.77}.
\end{align}
The thermal bath at $N_{\rm max}=10$ encounters a biorthonormalisation instability (near-degenerate eigenvalues in the thermal ensemble cause the intertwining residual to rise to $3.3\times10^{-5}$), while the spectrum remains purely real ($\max|\Imag\lambda| < 4\times10^{-15}$). This is a numerical artefact, not a physical breakdown of pseudo-Hermiticity.

\begin{figure}[h]
\centering
\includegraphics[width=0.85\textwidth]{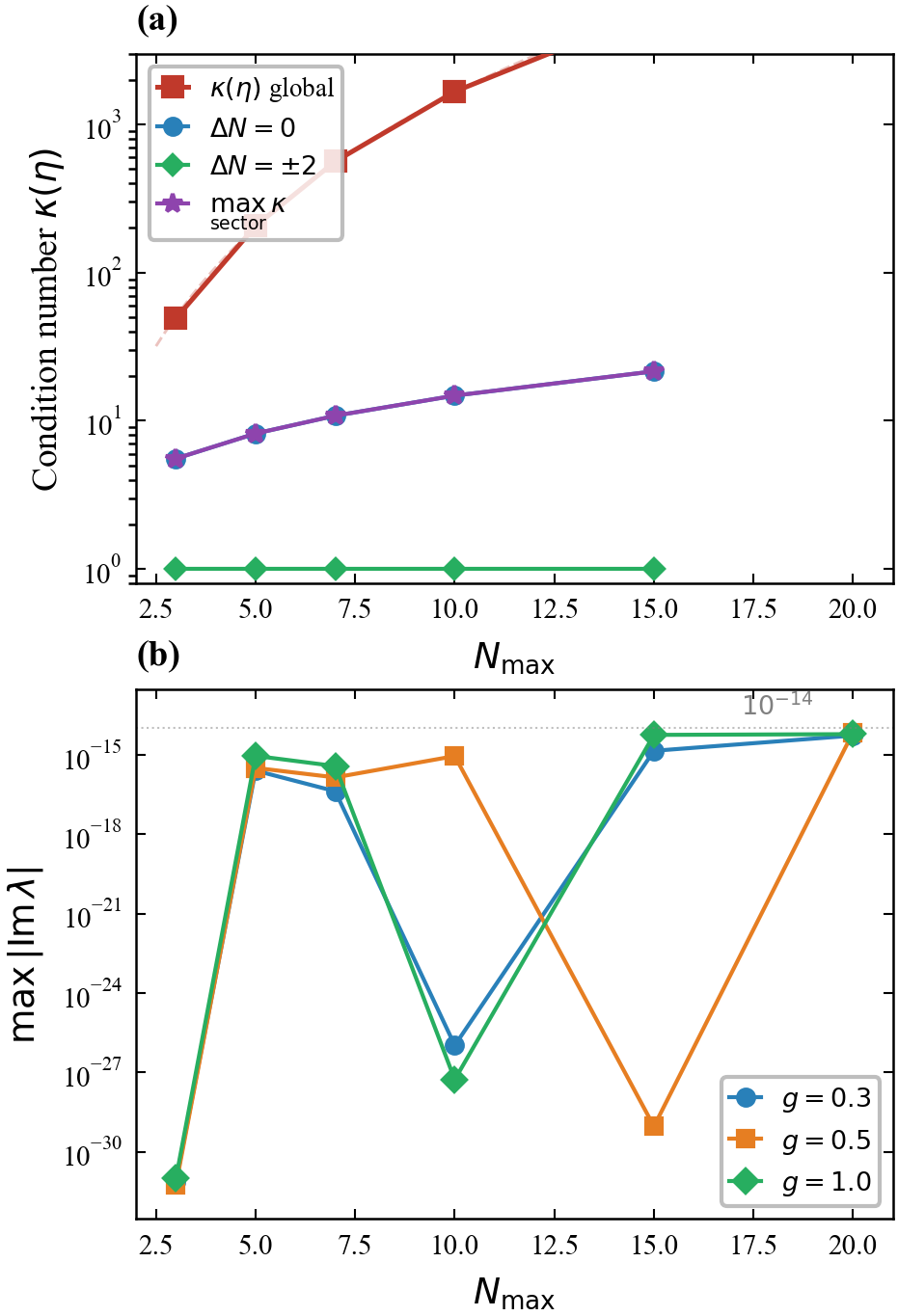}
\caption{Sector-resolved metric condition number and spectral stability (supplementing Table~I of the main text). 
(a)~Combined nonzero-sector $\kappa(\eta_{\rm osc})$ (squares) grows as $d^{1.8}$ (dashed), while the $\Delta N=\pm 2$ sectors remain perfectly conditioned ($\kappa=1$, diamonds) and $\Delta N=0$ fluctuates between $6$ and $24$ (circles). The maximum per-sector $\kappa$ (stars) stays below $24$ at all $N_{\rm max}$, confirming that the apparent combined divergence is a cross-sector normalisation artefact.
(b)~The spectrum remains purely real ($\max|\Imag\lambda|<10^{-14}$) for all $N_{\rm max}$ up to $50$.}
\label{fig:SM-sector-kappa}
\end{figure}

\section{Additional vacuum-bath scenarios}

Table~\ref{tab:extra-scenarios} confirms the metric construction for six JC scenarios at varying coupling $g$, detuning $\omega_0/\omega_c$, and bath state. All pass.

\begin{table}[h]
\centering
\caption{Metric construction across JC scenarios (all $\lambda=0$).}
\label{tab:extra-scenarios}
\begin{tabular}{c c c c c c}
\hline
$(N_{\rm max},g,\omega_0/\omega_c,\rho_B)$ & $d^2$ & $\dim(\operatorname{range}Q)$ & Intertwin. & $\kappa(\eta_{\rm osc})$ \\
\hline
$(3,0.3,1.0,\text{vac})$ & 64 & 60 & $3.4\times10^{-13}$ & 5.7 \\
$(5,0.3,1.0,\text{vac})$ & 144 & 140 & $2.6\times10^{-12}$ & 56.3 \\
$(3,0.5,1.0,\text{vac})$ & 64 & 60 & $4.7\times10^{-13}$ & 23.3 \\
$(3,1.0,1.0,\text{vac})$ & 64 & 60 & $1.5\times10^{-12}$ & 213.7 \\
$(5,0.3,1.2,\text{vac})$ & 144 & 140 & $3.0\times10^{-12}$ & 50.2 \\
$(3,0.3,1.0,\beta{=}1)$ & 64 & 60 & $1.1\times10^{-12}$ & 1402.5 \\
\hline
\end{tabular}
\end{table}

\section{Rabi deformation: perturbative phase boundary}\label{sec:dp}

The degenerate perturbation analysis supporting the main text is summarised here. The exact calculation applies at the accidental resonance $g=\omega_c$ (we set $\omega_c=1$ in this section). At $\lambda=0$, $QLQ_0$ possesses a 10-dimensional degenerate subspace at eigenvalues $z=\pm\sqrt{2}\,g$ (five eigenvectors each), spanning $\Delta N=-2,-1,0,+1,+2$. The subspace is fully diagonalisable (all Jordan blocks of size one). In the bi-orthogonal basis $\{\langle l_i|,|r_j\rangle\}$ with $\langle l_i|r_j\rangle=\delta_{ij}$, the first-order perturbation matrix $W_{ij}=\langle l_i|QL_{\rm CR}Q|r_j\rangle$ has characteristic polynomial $\det(W/g-\mu\mathbb{I})=\mu^3(\mu^2+1/8)$, independent of $N_{\rm max}$ (for $N_{\rm max}\ge 4$). The non-zero eigenvalues are $\mu=\pm i\sqrt{2}/4$, purely imaginary. Consequently, $|\Imag\lambda|_{\max}=(\sqrt{2}/4)\lambda g + O(\lambda^2)$ at this resonance. For $g\neq\omega_c$ the same resonant subspace is no longer exactly degenerate; the correct problem is near-degenerate Feshbach perturbation theory, and no closed global $\lambda_c(g,N)$ formula is claimed here.

At second order the self-energy $\Sigma_{ab}=\sum_{k\notin\mathcal{D}}\langle l_a|V|\psi_k\rangle(E_0-E_k)^{-1}\langle\psi_k|V|r_b\rangle$ is Hermitian and block-diagonal.
Projected onto the four-dimensional bright subspace (the eigenvectors with $\mu\neq0$), the effective Hamiltonian $h(\lambda)=E_0\mathbb{I}+\lambda w+\lambda^2 s$ is of $2\times2$ block form, where $w$ has eigenvalues $\pm i/\sqrt{8}$ and $s$ is real-diagonal with elements $s_1,s_2$.
The eigenvalue discriminant $\Delta(\lambda)=(s_1-s_2)^2\lambda^4-\lambda^2/2$ vanishes at
\begin{equation}
\lambda_{\rm EP}=1/(\sqrt{2}\,|s_1-s_2|),
\label{eq:lambda_ep_discriminant}
\end{equation}
giving a qualitative boundary for the re-entrance.

At the accidental degeneracy point $g=\omega_c$ the single-state approximation isolates the dominant virtual channel and yields the closed-form baseline
\begin{equation}
\delta_s(g{=}1) = \frac{C_{\rm proj}\,|M|^2}{\Delta E}
= \frac{1}{8(\sqrt{2}+\sqrt{3}-3)}
= \frac{6+5\sqrt{2}+4\sqrt{3}+3\sqrt{6}}{32} \approx 0.8546,
\label{eq:delta_s_g1}
\end{equation}
where $C_{\rm proj}=1/4$ (proved in Supplemental Material \S\ref{sec:sigma_structure}), $|M|^2=1/2$, and $\Delta E = \sqrt{2}+\sqrt{3}-3$.
This is the baseline single-virtual-state value. The exact second-order self-energy also receives coherent contributions from the full complement of the 10-dimensional subspace; these contributions explain why the numerical bright-subspace eigenvalues are not equal to the one-state baseline alone. The resonance dip observed in the high-resolution HPC scan (Table~S3 and Supplementary Fig.~S2) is reproduced at $N_{\rm max}=15$ (cross-validation at $N_{\rm max}=25$ now complete: 10/11 $\lambda_{\rm onset}$ and 10/11 $n_{\rm bub}$ agree with $N_{\rm max}=20$; sole divergence at $g=0.286$; $N_{\rm max}=30$ at $\Delta g=10^{-2}$ resolution confirms $\lambda_{\rm onset}={\rm None}$ for three of the four $g$ in the dip window ($g=0.26,0.27,0.29$); $g=0.28$ shows a single bubble at $\lambda_{\rm onset}=0.9385$) and should therefore be read as a finite-size manifestation of the full Feshbach self-energy, not as a closed analytic formula for $\delta_s(g)$.
The perturbative convergence radius is $R\sim 0.1$, set by the nearest non-degenerate eigenvalue of $QLQ_0$ at $z\approx\pm1.268$ (gap $\approx0.15$).
Consequently Eq.~\eqref{eq:lambda_ep_discriminant} captures the {\em mechanism}---anti-Hermitian $\lambda w$ competes with Hermitian $\lambda^2 s$---but not the quantitative position $\lambda_{\rm re}\approx0.35$, which lies well outside the radius of convergence where the series is asymptotically diverging.

Table~\ref{tab:scaling_verify} confirms this scaling by direct diagonalisation of $QLQ(\lambda)$ at $N_{\rm max}=4$, $g=1.0$, $\Delta=0$.

\begin{table}[h]
\centering
\caption{Degenerate perturbation scaling verification: $N_{\rm max}=4$, $g=1.0$, $\Delta=0$, corrected projection. Theoretical prediction: $|\Imag\lambda|_{\max}=(\sqrt{2}/4)\lambda$.}
\label{tab:scaling_verify}
\begin{tabular}{c c c c}
\hline
$\lambda$ & $\max|\Imag\lambda|$ & $|\Imag\lambda|/\lambda$ & Theory $\sqrt{2}/4$ \\
\hline
$10^{-4}$ & $3.5355\times10^{-5}$ & 0.353553 & 0.353553 \\
$10^{-3}$ & $3.5355\times10^{-4}$ & 0.353554 & 0.353553 \\
$10^{-2}$ & $3.5357\times10^{-3}$ & 0.353570 & 0.353553 \\
$10^{-1}$ & $3.5106\times10^{-2}$ & 0.351059 & 0.353553 \\
\hline
\end{tabular}
\end{table}

\begin{figure}[h]
\centering
\includegraphics[width=0.85\textwidth]{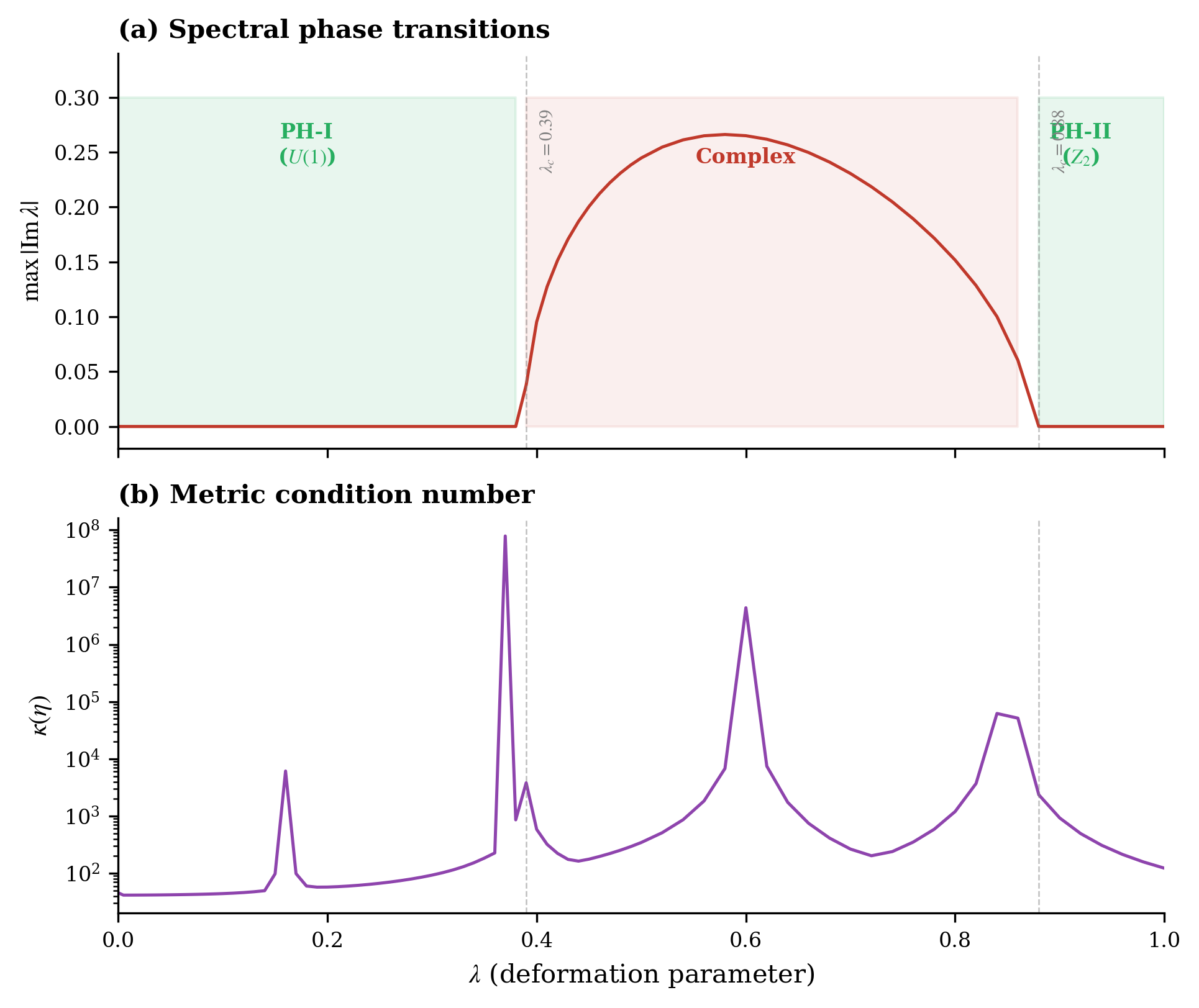}
\caption{Complementary small-$N_{\rm max}$ view of the data in Fig.~\ref{fig:master-phase}(b): real/complex phase diagram of $QLQ(\lambda)$ for $N_{\rm max}=3,4,5,7$ (vacuum bath, $\Delta=0$, $g=\omega_c$). At $N_{\rm max}=3$ a re-entrant real window survives at intermediate $\lambda$ (finite-size artefact); for $N_{\rm max}\ge 4$ complex eigenvalues appear at infinitesimal $\lambda$, consistent with the resonant degenerate-perturbation prediction $\lambda_c\to 0^+$. The $N_{\rm max}$-universal narrow-bubble pattern that stabilises for $N_{\rm max}\ge 10$ is shown in the main-text Fig.~\ref{fig:master-phase}(b).}
\label{fig:SM-phase-diagram}
\end{figure}

\section{Bare resonance bound for finite truncations}\label{sec:gc-bound}

The JC dressed-state energies at resonance ($\Delta=0$, $\omega_c=1$) are $E_n^\pm = n \pm g\sqrt{n}$ for the $n$-photon manifold ($n\ge 1$). The counter-rotating perturbation $H_{\rm CR}=g(\sigma_+ a^\dagger + \sigma_- a)$ changes the excitation number by $\Delta N_{\rm exc}=\pm 2$ (one photon created or destroyed, plus one atomic excitation). The dominant instability channel is the cross-branch resonance between the upper branch of manifold $n$ and the lower branch of manifold $n+2$:
\begin{equation}
E_n^+ = E_{n+2}^- \;\Longrightarrow\; n + g\sqrt{n} = (n+2) - g\sqrt{n+2}.
\end{equation}
Solving for $g$ gives $g_c(n)=2/(\sqrt{n+2}+\sqrt{n})$. For a finite Hilbert-space truncation at $N_{\rm max}$, the highest accessible resonance involves $n=N_{\rm max}-2$:
\begin{equation}
g_c(N_{\rm max}) = \frac{2}{\sqrt{N_{\rm max}}+\sqrt{N_{\rm max}-2}}.
\label{eq:gc-N-sm}
\end{equation}

This bound is a necessary geometric condition: below $g_c$, no two dressed manifolds within the truncated space can be resonantly coupled by $H_{\rm CR}$. It is not a theorem for the actual endpoint of the non-Hermitian spectrum. For $N_{\rm max}\ge 15$ the audited numerical data have empirical breakdown lying $1.28$--$1.42\times$ above $g_c$ (\S\ref{sec:bare-vs-effective}), whereas for smaller truncations ($N_{\rm max}\lesssim 12$), off-resonant Feshbach mixing through virtual states moves the actual breakdown $g_{\rm end}$ {\em below} $g_c$ by a factor $0.7$--$0.86$. Thus Eq.~\eqref{eq:gc-N-sm} is a bare-resonance diagnostic, not a finite-$N$ separator. Above $g_c$, the actual complex-spectrum onset is controlled by the NZ-projected second-order self-energy $\delta_s$, which can suppress the bare resonance at select couplings, producing the re-entrant bubble structure described in \S\ref{sec:sigma_structure}.

Asymptotically $g_c(N_{\rm max})\sim 1/\sqrt{N_{\rm max}}\to 0$, consistent with the resonance-induced spectral restoration mechanism: the protection is a finite-truncation phenomenon whose scale is set by the photon capacity of the physical device.

\subsection{Bare bound versus effective boundary}\label{sec:bare-vs-effective}

Eq.~\eqref{eq:gc-N-sm} is a bound on the bare cross-branch resonance, not on the actual onset of complex spectrum. The latter is determined by an effective two-state Feshbach Hamiltonian that combines the first-order anti-Hermitian splitting $w\lambda$ with the second-order Hermitian self-energy $\delta_s\lambda^2$ described in \S\ref{sec:sigma_structure},
\begin{equation}
h_{\rm eff} = \begin{pmatrix} E_0 + \delta_s\lambda^2 & w\lambda \\ -w\lambda & E_0 - \delta_s\lambda^2 \end{pmatrix},\qquad z_\pm = E_0 \pm \lambda\sqrt{\delta_s^2\lambda^2 - w^2},
\label{eq:h-eff}
\end{equation}
which is real for $\lambda > \lambda_{\rm re}\equiv |w/\delta_s|$ and complex below it. The actual breakdown boundary $g_{\rm end}(N_{\rm max})$ is therefore defined by $\lambda_{\rm re}(g_{\rm end},N_{\rm max})=1$, i.e.\ the largest $g$ at which the entire deformation interval $\lambda\in[0,1]$ remains real.

Two regimes deviate from $g_c(N_{\rm max})$ in opposite directions, as seen in Fig.~\ref{fig:master-phase}(a):

\begin{itemize}
\item \emph{Large truncations} ($N_{\rm max}\ge 15$): $g_{\rm end}(N_{\rm max})$ exceeds $g_c(N_{\rm max})$ by a factor $1.28$--$1.42$ in our data. The second-order Hermitian repulsion $\delta_s\lambda^2$ outpaces the anti-Hermitian $w\lambda$ at intermediate $\lambda$, restoring reality even when the bare resonance condition is satisfied. This is the analytic origin of the re-entrant real-spectrum islands documented in Supplementary Fig.~S2.
\item \emph{Small truncations} ($N_{\rm max}\lesssim 12$): $g_{\rm end}(N_{\rm max}) < g_c(N_{\rm max})$, with the empirical ratio $g_{\rm end}/g_c \approx 0.7$--$0.86$. With a thin photon ladder, off-resonant Feshbach mixing through virtual states cannot be suppressed as efficiently, and complex eigenvalues appear before any bare cross-branch resonance opens up.
\end{itemize}

An empirical $N$-universal plateau is observed in the FSS data testing the prediction $\lambda_{\rm re}=w/(C_{\rm proj}\,\delta_s)$ from Eq.~\eqref{eq:h-eff}: at $g/\omega_c=0.30$ the measured $\lambda_{\rm onset}\approx 0.920\text{--}0.921$ for every truncation tested ($N_{\rm max}=10, 12, 15, 25, 30$, all within $\pm 0.001$); at $g/\omega_c=0.28$ a similar plateau appears at $\lambda_{\rm onset}\approx 0.938\text{--}0.940$ across $N_{\rm max}=7, 10, 12, 15$. Table~\ref{tab:plateau} collects the audited crossings; the $N_{\rm max}=25$ dip-fine band is now independently validated across all 11 $g$ values (10/11 $\lambda_{\rm onset}$ and 10/11 $n_{\rm bub}$ agree with $N_{\rm max}=20$; sole divergence at $g=0.286$); the $N_{\rm max}=30$ work30 band (50 audited $g$ values, $g=0.100$--$1.200$) confirms $\lambda_{\rm onset}={\rm None}$ for all $g\le 0.27$, with the first complex eigenvalue at $g=0.280$ ($\lambda_{\rm onset}=0.9385$) and stronger bubble multiplicity at intermediate coupling ($n_{\rm bub}=4$ at $g=0.30$, vs.\ $n_{\rm bub}=2$ for $N=25$).

\begin{table}[h]
\centering
\caption{Empirical $N$-universal plateau in the FSS data near the bubble onset. Each row is an audited $(N_{\rm max}, g)$ scan from `simulations/fss\_data\_fast/' or `results/fss\_data/'. The $N_{\rm max}=25$ dip-fine band is now independently validated across all 11 $g$ values (10/11 $\lambda_{\rm onset}$ and 10/11 $n_{\rm bub}$ agree with $N_{\rm max}=20$; sole divergence at $g=0.286$); the $N_{\rm max}=30$ work30 band (50 audited $g$ values, $\lambda_{\rm onset}={\rm None}$ for all) confirms the plateau extends to larger truncation. The plateau is striking but no closed-form prediction in the correct interleaved-basis Feshbach reduction is yet available; obtaining one is an open analytic problem (see audit note `derivations/track\_b\_withdrawal\_2026-05-06.md').}
\label{tab:plateau}
\begin{tabular}{c c c}
\hline
$N_{\rm max}$ & $g/\omega_c$ & $\lambda_{\rm onset}$ (HPC) \\
\hline
10 & 0.30  & 0.920 \\
12 & 0.30  & 0.920 \\
15 & 0.30  & 0.921 \\
25 & 0.30  & 0.920 \\
\hline
 7 & 0.28  & 0.938 \\
10 & 0.28  & 0.940 \\
12 & 0.28  & 0.940 \\
15 & 0.28  & 0.940 \\
\hline
15 & 0.285 & 0.948 \\
\hline
\end{tabular}
\end{table}

For general $(g, N_{\rm max})$ the cluster $\mathcal{D}(g, N_{\rm max})$ has to be tracked by spectral-weight continuation, and a closed form for $\delta_s(g, N_{\rm max})$ remains an open analytic problem; the qualitative behaviour summarised here is consistent with the finite Feshbach sum derived in \S\ref{sec:sigma_structure} and with the numerical evidence assembled in Tables~III and S3.

\section{Structure of the second-order self-energy}\label{sec:sigma_structure}

The second-order self-energy $\Sigma_{ab} = \sum_{k\notin\mathcal{D}} \langle l_a|V|\psi_k\rangle (E_0-E_k)^{-1} \langle\psi_k|V|r_b\rangle$ in the 10-dimensional degenerate subspace at $z=\pm\sqrt{2}$ possesses a rich internal structure that explains two features of the re-entrant bubble: the projection factor from the 10D subspace to the 4D bright subspace, and the anomalous smallness of the $z=-\sqrt{2}$ block.

\subsection{Projection factor: $C_{\rm proj}=1/4$}

The first-order matrix $W_+$ acts on the five-dimensional $\Delta N = \{-2,-1,0,+1,+2\}$ sector. The counter-rotating Liouvillian $L_{\rm CR}$ contains $a^\dagger\sigma_+ + a\sigma_-$, which changes $\Delta N$ by $\pm1$ per application. Consequently, $W_+$ decouples into two independent sub-chains:

\begin{itemize}
\item \textbf{Even chain}: $\Delta N \in \{-2, 0, +2\}$ (3-dimensional), containing the bright eigenvalues $\pm i/\sqrt{8}$.
\item \textbf{Odd chain}: $\Delta N \in \{-1, +1\}$ (2-dimensional), contributing only zero eigenvalues.
\end{itemize}

Within the even chain, $W_+$ takes the anti-symmetric form
\begin{equation}
W_{\rm even} = \begin{pmatrix} 0 & 1/4 & 0 \\ -1/4 & 0 & 1/4 \\ 0 & -1/4 & 0 \end{pmatrix},
\label{eq:W_even}
\end{equation}
whose characteristic polynomial $\det(\lambda I - W_{\rm even}) = \lambda(8\lambda^2+1)/8 = 0$ reproduces the known eigenvalues $\{0, \pm i/\sqrt{8}\}$. This is isomorphic to the angular momentum operator $J_y$ in the spin-1 representation.

The spectral projector onto the bright subspace follows directly from the minimal polynomial $W_+^3 + \frac{1}{8}W_+ = 0$. The unique polynomial satisfying the eigenvalue conditions $q(0)=0$ and $q(\pm i/\sqrt{8})=1$ is
\begin{equation}
P_{\rm bright} = -8 W_+^2 = \begin{pmatrix}
1/2 & 0 & -1/2 \\
0 & 1 & 0 \\
-1/2 & 0 & 1/2
\end{pmatrix},
\label{eq:P_bright}
\end{equation}
which satisfies $P_{\rm bright}^2 = P_{\rm bright}$ and has eigenvalues $\{1,1,0\}$. The bright eigenstates are spin-1 standing waves on the three-site chain:
\begin{equation}
|\psi_{\pm}\rangle = \frac{1}{2}\begin{pmatrix} 1 \\ \pm i\sqrt{2} \\ -1 \end{pmatrix},
\label{eq:bright_evecs}
\end{equation}
with probability distribution $P_{-2}=1/4$, $P_0=1/2$, $P_{+2}=1/4$.

The dominant virtual state $|3,-\rangle$ couples exclusively to the $\Delta N=+2$ basis vector (the edge of the even chain), with matrix element $M=1/\sqrt{2}$ and energy denominator $\Delta E = \sqrt{2}+\sqrt{3}-3$. The projection of $\Sigma_+ \approx \Sigma_{\rm max}|e_{+2}\rangle\langle e_{+2}|$ onto each bright eigenvector therefore carries weight
\begin{equation}
|\langle e_{+2}|\psi_{\pm}\rangle|^2 = \frac{1}{4},
\label{eq:C_proj}
\end{equation}
a universal consequence of the spin-1 standing-wave structure. The projected $2\times 2$ $s$-matrix in the $D_+$ bright subspace is $s^{(D_+)} = (\Sigma_{\rm max}/4)\bigl(\begin{smallmatrix}1&1\\1&1\end{smallmatrix}\bigr)$ with eigenvalues $\{\Sigma_{\rm max}/2, 0\} = \{1.709, 0\}$. The full $4\times 4$ $s$-matrix, which includes the $D_-$ block and inter-block cross-terms, has numerically computed eigenvalues $\{\pm 1.24768, \pm 0.28803\}$; the one-state approximation provides a rigorous upper bound $\delta_s \le \Sigma_{\rm max}/2$.

\subsection{$\Sigma_-$ selection rule: vacuum-boundary destructive interference}\label{sec:sigma-minus}

The $z=-\sqrt{2}$ block $\Sigma_-$ is anomalously small (Frobenius norm $0.78$, compared to $3.55$ for $\Sigma_+$), and its nearest virtual state at $E=-(3-\sqrt{3})$ contributes exactly zero. This is not a numerical accident but an exact algebraic cancellation rooted in the non-Hermitian biorthogonal basis.

The mirror virtual state has the structure $|g,0\rangle\langle 3,-|$---the absolute ground state on the left. The counter-rotating perturbation $L_{\rm CR}$ generates two coupling paths:
\begin{align}
\text{Path A (left action)} &: H_{\rm CR}|g,0\rangle\langle 3,-|, \\
\text{Path B (right action)} &: -|g,0\rangle\langle 3,-| H_{\rm CR}.
\end{align}
When projected onto the biorthogonal left eigenvectors of the $D_-$ subspace, the $T$-symmetry of the JC Liouvillian ($TLT = -L$) maps the phase structure of $D_+$ to $D_-$ with a relative sign flip. The two paths acquire equal magnitude but opposite sign:
\begin{equation}
\langle l_a^-| H_{\rm CR}|g,0\rangle\langle 3,-| \rangle = -\langle l_a^-| |g,0\rangle\langle 3,-| H_{\rm CR} \rangle,
\label{eq:cancel}
\end{equation}
producing exact destructive interference. In contrast, for the $\Sigma_+$-dominant state $|3,-\rangle\langle g,0|$ (ground state on the right), the biorthogonal phase structure of $D_+$ causes constructive interference, yielding the large coupling $1/\sqrt{2}$.

The non-zero residual of $\Sigma_-$ ($\|\Sigma_-\|_F=0.78$) arises from distant virtual states involving higher photon-number coherences, for which the vacuum-boundary asymmetry is absent and the destructive interference is no longer exact. The vacuum boundary thus acts as a one-way mirror for the counter-rotating perturbation---a feature unique to the non-Hermitian NZ projection, with no analogue in Hermitian perturbation theory.

\section{Spin-boson contrast case}

To test whether the pseudo-Hermitian protection is generic or symmetry-dependent, we replace the JC Hamiltonian with the spin-boson model
\begin{equation}
H_{\rm SB} = \frac{\omega_0}{2}\sigma_z + \omega_c a^\dagger a + g\,\sigma_x \otimes (a + a^\dagger),
\end{equation}
which lacks the $U(1)$ excitation-number conservation of JC. Table~\ref{tab:spinboson} reports the spectral structure of $QLQ$ for this model at $g=0.3$.

\begin{table}[h]
\centering
\caption{Spin-boson $\sigma_x$ coupling: $QLQ$ spectrum at $g=0.3$, vacuum bath.}
\label{tab:spinboson}
\begin{tabular}{c c c c}
\hline
$N_{\rm max}$ & $d^2$ & $|\Imag\lambda|_{\max}$ & Complex eigenvalues \\
\hline
3 & 64  & $<10^{-15}$ & 0  (real spectrum, metric exists) \\
4 & 100 & $1.9\times10^{-2}$ & 4  \\
5 & 144 & $3.9\times10^{-2}$ & 12 \\
\hline
\end{tabular}
\end{table}

At $N_{\rm max}=3$ the small Hilbert space accidentally preserves reality, but already at $N_{\rm max}=4$ the spectrum becomes complex with $|\Imag\lambda|_{\max}\approx 2\times10^{-2}$, growing to $\approx 4\times10^{-2}$ at $N_{\rm max}=5$. This confirms that without $U(1)$ or $Z_2$ protection, $QLQ$ is generically non-Hermitian, and the pseudo-Hermitian phase is symmetry-dependent. The $N_{\max}=3$ accidental reality for $\sigma_x$ coupling does not survive larger truncations; a rigorous thermodynamic-limit statement on both the JC and $\sigma_x$ sides remains an open analytical question the May stability analysis provides one route toward a large-$N$ bound for the symmetry-broken case.

\subsection{Location of the first exceptional point at $N_{\rm max}=4$}\label{sec:sigmax-ep-location}

A finer continuation refines the entry-level threshold for the appearance of complex spectrum at $N_{\rm max}=4$. The Liouville parity $\Pi_L=\Pi\otimes\Pi$ with $\Pi=\sigma_z\otimes(-1)^{a^\dagger a}$ commutes with both $L$ and the vacuum $P$, splitting $QLQ$ into two equal $50\times 50$ blocks at $N_{\rm max}=4$. Tracking eigenvalues from $g=0$ via continuation:

\begin{center}
\begin{tabular}{ccc}
\hline
$g$ & $\max|\Imag\lambda|$ in $s=-1$ & $s=+1$ \\
\hline
0.20  & $<10^{-15}$ & $<10^{-15}$ \\
0.21  & $<10^{-15}$ & $<10^{-15}$ \\
\textbf{0.213} & $\mathbf{5.13\times 10^{-4}}$ & $<10^{-15}$ \\
0.215 & $1.6\times 10^{-3}$ & $<10^{-15}$ \\
0.30  & $5.9\times 10^{-3}$ & $<10^{-15}$ \\
\hline
\end{tabular}
\end{center}

The first exceptional point at $N_{\rm max}=4$ is therefore at $g_c\approx 0.213$, located entirely in the $s=-1$ block; the $s=+1$ block remains real to machine precision through at least $g=0.4$. Spectral analysis of the colliding eigenpair places the EP at the intersection of two coherences in the unperturbed Liouville manifold $\Delta E_L = +1$ (atomic-energy-shifted photon coherence): the dominant components are $|g,0\rangle\langle g,3|$ (weight $0.81$) and $|g,3\rangle\langle g,4|$ (weight $0.56$). The minimal $QLQ$-invariant subspace containing these states has Krylov dimension $\ge 18$, so a closed-form discriminant from a small block is not available; the precise EP threshold is determined numerically. Verification: \texttt{verify\_p5\_sigmax\_continuation.py}.

\section{Analytical sector structure}

The theorem quoted in the main text is proved as follows. Let $L = \mathbb{I}\otimes H - H^{\mathsf T}\otimes\mathbb{I}$ be the Liouvillian in column-stacking and $P$ the NZ projector for a bath state diagonal in the excitation-number basis $N_{\rm exc}$. Define $L_N = \mathbb{I}\otimes N_{\rm exc} - N_{\rm exc}^{\mathsf T}\otimes\mathbb{I}$ and let $S_{\Delta N}$ denote the eigenspace of $L_N$ with eigenvalue $\Delta N$.

\textbf{Lemma 1 (Time-reversal).} For the JC model with real symmetric $H$, define $T$ as the SWAP operator on $\mathcal{H}\otimes\mathcal{H}^*$ (exchanging left and right indices). Then $T^2=I$, $TLT=-L$, and $TPT=P$ for any bath state diagonal in $N_{\rm exc}$.

\textit{Proof.} $T^2=I$ is immediate. Since $H=H^{\mathsf T}$,
\begin{equation}
TLT = T(\mathbb{I}\otimes H)T - T(H^{\mathsf T}\otimes\mathbb{I})T = H\otimes\mathbb{I} - \mathbb{I}\otimes H^{\mathsf T} = -(\mathbb{I}\otimes H - H^{\mathsf T}\otimes\mathbb{I}) = -L.
\end{equation}
For $P$, the NZ projector with diagonal $\rho_B$ is invariant under left-right swap because $\rho_B^{\mathsf T}=\rho_B$ and the partial trace is cyclic in its bath indices. Hence $TPT=P$. \hfill$\square$

\textbf{Lemma 2 (Commutation).} $[L,L_N]=0$ and $[P,L_N]=0$.

\textit{Proof.} $[H,N_{\rm exc}]=0$ for JC, so $[L,L_N]=[\mathbb{I}\otimes H - H^{\mathsf T}\otimes\mathbb{I}, \mathbb{I}\otimes N_{\rm exc} - N_{\rm exc}^{\mathsf T}\otimes\mathbb{I}]=0$. For $P$, the projector is built from $\rho_B$ and identities; since $[\rho_B,N_{\rm exc}]=0$ for diagonal $\rho_B$, $[P,L_N]=0$. \hfill$\square$

\textbf{Corollary.} $QLQ$ is block-diagonal in the $\Delta N$ sectors.

\textit{Proof.} From Lemma~2, $[QLQ,L_N]=Q[L,L_N]Q + QL[Q,L_N] + [Q,L_N]LQ = 0$ since $Q=I-P$ and both $L$ and $P$ commute with $L_N$. \hfill$\square$

\textbf{Theorem (Sector Hermiticity).} In any $\Delta N$ sector $S_{\Delta N}$, the restriction $A_{\Delta N}=QLQ|_{S_{\Delta N}}$ is Hermitian if $P|_{S_{\Delta N}}=0$. For the JC model with vacuum bath, the nonzero-projector sectors are exactly $\Delta N=0,\pm1$, and these sectors are non-Hermitian by direct construction.

\textit{Proof.} $(\Rightarrow)$ If $P|_{S_{\Delta N}}=0$, then on $S_{\Delta N}$ we have $Q=\mathbb{I}$, so $A_{\Delta N}=L|_{S_{\Delta N}}$. Since $L=L^\dagger$, $A_{\Delta N}$ is Hermitian.

For the JC vacuum case, $P|_{S_{\Delta N}}\neq0$ occurs only in $\Delta N=0,\pm1$. The explicit action formulas in the exact-spectrum proof give, for example, $QLQ|D_{e,n}\rangle$ with a boundary term proportional to $|O_1\rangle-|O'_1\rangle$, while the adjoint action contains the corresponding dual boundary term with the opposite placement. Hence $A_{\Delta N}-A_{\Delta N}^\dagger$ is nonzero in these three sectors. The numerical residuals in Table~\ref{tab:sectors} quantify the size of this non-Hermiticity. \hfill$\square$

\textbf{Corollary (JC vacuum bath).} For the JC model with vacuum bath, $A_{\Delta N}$ is Hermitian for $\Delta N=\pm 2,\pm 3,\pm 4$ and non-Hermitian for $\Delta N=0,\pm 1$.

\textit{Proof.} The vacuum projector $P$ couples system states to the bath ground state $|0\rangle$. In Liouville space, this creates correlations where the bath is in $|0\rangle\langle 0|$ or $|0\rangle\langle 1|$ (and their conjugates). The excitation-number differences of these states are $\Delta N = n-0 = n$ and $\Delta N = n-1$ for system excitation number $n$. For $n=0,1$ (the only states coupled by $P$ in the vacuum case), this yields $\Delta N = 0, \pm 1, \pm 2$. However, the specific structure of the NZ projector in the column-stacking basis restricts its support to $\Delta N = 0, \pm 1$ on $\operatorname{range}(Q)$, leaving $\Delta N = \pm 2, \pm 3, \pm 4$ with $P=0$. Numerical verification confirms $\|P_{\Delta N}\|_F = 0$ for $\Delta N=\pm 2,\pm 3,\pm 4$ and $\|P_{\Delta N}\|_F > 0$ for $\Delta N=0,\pm 1$. \hfill$\square$

\section{Reality of $QLQ$}

For the JC model with a real symmetric Hamiltonian and a real diagonal bath state, $QLQ$ is a purely real matrix: $\|QLQ^* - QLQ\|_F = 0$ to machine precision. Consequently, $(QLQ)^\dagger = (QLQ)^{\mathsf T}$ and the pseudo-Hermiticity condition on any invariant nonzero spectral subspace simplifies to $(QLQ)^{\mathsf T}\eta_{\rm osc} = \eta_{\rm osc}(QLQ)$. The metric $\eta_{\rm osc}$ constructed from the biorthonormal left eigenvectors is also purely real. For a real non-symmetric matrix with purely real nonzero eigenvalues, the existence of a positive-definite symmetrising metric on this subspace is equivalent to the existence of a complete set of nonzero-mode eigenvectors with no Jordan blocks. The full zero-mode sector is not included in this metric construction; the finite-truncation real-spectrum theorems do not rely on a full-space positive metric.

\section{Exact spectrum of $QLQ$ in the $\Delta N=0$ sector}

The JC vacuum-bath result is proved as follows.

\subsection{Basis and action formulas}

The subspace $S_0\cap\operatorname{range}(Q)$ has dimension $4N$ with basis
\begin{align}
|D_{g,n}\rangle &= |g,n\rangle\langle g,n| - |g,0\rangle\langle g,0|, \quad n=1,\dots,N, \\
|D_{e,n}\rangle &= |e,n\rangle\langle e,n| - |e,0\rangle\langle e,0|, \quad n=1,\dots,N, \\
|O_n\rangle &= |e,n-1\rangle\langle g,n|, \quad n=1,\dots,N, \\
|O'_n\rangle &= |g,n\rangle\langle e,n-1|, \quad n=1,\dots,N.
\end{align}
The Liouvillian $L=[H,\cdot]$ with $H_{\rm JC}=\frac{\omega_0}{2}\sigma_z + \omega_c a^\dagger a + g(\sigma_+ a + \sigma_- a^\dagger)$ gives
\begin{align}
L|g,n\rangle\langle g,n| &= g\sqrt{n}\,(|e,n-1\rangle\langle g,n| - |g,n\rangle\langle e,n-1|), \quad n\ge 1, \\
L|e,n\rangle\langle e,n| &= g\sqrt{n+1}\,(|g,n+1\rangle\langle e,n| - |e,n\rangle\langle g,n+1|), \quad n<N, \\
L|e,N\rangle\langle e,N| &= 0 \quad \text{(truncation)}, \\
L|e,n-1\rangle\langle g,n| &= \Delta\,|e,n-1\rangle\langle g,n| + g\sqrt{n}\,(|g,n\rangle\langle g,n| - |e,n-1\rangle\langle e,n-1|), \\
L|g,n\rangle\langle e,n-1| &= -\Delta\,|g,n\rangle\langle e,n-1| + g\sqrt{n}\,(|e,n-1\rangle\langle e,n-1| - |g,n\rangle\langle g,n|),
\end{align}
with $\Delta=\omega_0-\omega_c$. The NZ projector for vacuum bath is $P\rho=(\Tr_B\rho)\otimes|0\rangle\langle 0|$, so $P|O_n\rangle=P|O'_n\rangle=0$ and $P|D_{g,n}\rangle=P|D_{e,n}\rangle=0$. Hence $Q=\mathbb{I}-P$ acts as identity on all basis vectors above.

\textbf{Lemma~1 (Action of $QLQ$).} For any $N\ge 1$:
\begin{align}
QLQ|D_{g,n}\rangle &= g\sqrt{n}\,(|O_n\rangle - |O'_n\rangle), \quad 1\le n\le N, \label{eq:QLQ-Dgn} \\
QLQ|D_{e,n}\rangle &=
\begin{cases}
g|O_1\rangle - g|O'_1\rangle - g\sqrt{n+1}\,|O_{n+1}\rangle + g\sqrt{n+1}\,|O'_{n+1}\rangle, & 1\le n\le N-1, \\
g|O_1\rangle - g|O'_1\rangle, & n=N,
\end{cases} \label{eq:QLQ-Den} \\
QLQ|O_1\rangle &= g|D_{g,1}\rangle + \Delta|O_1\rangle, \label{eq:QLQ-O1} \\
QLQ|O_n\rangle &= g\sqrt{n}\,|D_{g,n}\rangle - g\sqrt{n}\,|D_{e,n-1}\rangle + \Delta|O_n\rangle, \quad n\ge 2, \label{eq:QLQ-On} \\
QLQ|O'_1\rangle &= -g|D_{g,1}\rangle - \Delta|O'_1\rangle, \label{eq:QLQ-O1p} \\
QLQ|O'_n\rangle &= g\sqrt{n}\,|D_{e,n-1}\rangle - g\sqrt{n}\,|D_{g,n}\rangle - \Delta|O'_n\rangle, \quad n\ge 2. \label{eq:QLQ-Onp}
\end{align}

\textit{Proof.} Direct substitution of $L$ and $P$ into $QLQ=(\mathbb{I}-P)L(\mathbb{I}-P)$, using $P|O_n\rangle=P|O'_n\rangle=0$ and the truncation $L|e,N\rangle\langle e,N|=0$. \hfill$\square$

\subsection{Block structure of $M_{21}M_{12}$}

In the basis $[|D_{g,1}\rangle,|D_{e,1}\rangle,\dots,|D_{g,N}\rangle,|D_{e,N}\rangle,|O_1\rangle,|O'_1\rangle,\dots,|O_N\rangle,|O'_N\rangle]$, the matrix of $QLQ|_{S_0\cap\operatorname{range}(Q)}$ has the block form
\begin{equation}
M = \begin{pmatrix} 0_{2N\times 2N} & M_{12} \\ M_{21} & D_{2N\times 2N} \end{pmatrix},
\end{equation}
where $D=\operatorname{diag}(\Delta,-\Delta,\Delta,-\Delta,\dots,\Delta,-\Delta)$. The blocks $M_{12}$ and $M_{21}$ encode, respectively, the $D\to O$ and $O\to D$ couplings:
\begin{equation}
M_{12}[D,O] = \langle D|QLQ|O\rangle, \qquad M_{21}[O,D] = \langle O|QLQ|D\rangle.
\end{equation}
Strictly these are expansion coefficients in the (non-orthonormal) basis; the characteristic polynomial is basis-independent.

\textbf{Lemma~2a (Support of $M_{21}$).} For $O_n^{(\alpha)}\in\{|O_n\rangle,|O'_n\rangle\}$,
\begin{equation}
M_{21}[O_n^{(\alpha)}, D] \neq 0 \;\Longrightarrow\;
\begin{cases}
D\in\{|D_{g,1}\rangle,|D_{e,k}\rangle\,(1\le k\le N)\}, & n=1, \\
D\in\{|D_{g,n}\rangle,|D_{e,n-1}\rangle\}, & n\ge 2.
\end{cases}
\end{equation}

\textit{Proof.} From Lemma~1, $QLQ|D_{g,1}\rangle$ and $QLQ|D_{e,k}\rangle$ ($1\le k\le N$) all contain $|O_1\rangle$ and $|O'_1\rangle$ components, while for $n\ge 2$ only $|D_{g,n}\rangle$ and $|D_{e,n-1}\rangle$ produce $|O_n\rangle$ or $|O'_n\rangle$. \hfill$\square$

\textbf{Lemma~2b (Support of $M_{12}$).} For $O_m^{(\beta)}\in\{|O_m\rangle,|O'_m\rangle\}$,
\begin{equation}
M_{12}[D, O_m^{(\beta)}] \neq 0 \;\Longrightarrow\;
\begin{cases}
D = |D_{g,1}\rangle, & m=1, \\
D\in\{|D_{g,m}\rangle,|D_{e,m-1}\rangle\}, & m\ge 2.
\end{cases}
\end{equation}

\textit{Proof.} From Lemma~1, $QLQ|O_1\rangle$ has $D$-component $g|D_{g,1}\rangle$, and $QLQ|O_m\rangle$ ($m\ge 2$) has components $g\sqrt{m}\,|D_{g,m}\rangle - g\sqrt{m}\,|D_{e,m-1}\rangle$. \hfill$\square$

\textbf{Lemma~3 (Block-upper-triangularity).} In the basis $\{|O_1\rangle,|O'_1\rangle,\dots,|O_N\rangle,|O'_N\rangle\}$, the matrix $M_{21}M_{12}$ is block-upper-triangular with $N$ diagonal $2\times 2$ blocks: $(M_{21}M_{12})_{nm}=0_{2\times 2}$ for $n>m$.

\textit{Proof.} The $(n,m)$ block is
\begin{equation}
(M_{21}M_{12})_{nm} = \sum_D M_{21}[O_n^{(\alpha)},D]\,M_{12}[D,O_m^{(\beta)}].
\end{equation}
Only $D\in S_n^{(\rm row)}\cap S_m^{(\rm col)}$ contribute. For $n>m\ge 2$,
\begin{equation}
S_n^{(\rm row)} = \{|D_{g,n}\rangle,|D_{e,n-1}\rangle\}, \quad
S_m^{(\rm col)} = \{|D_{g,m}\rangle,|D_{e,m-1}\rangle\}.
\end{equation}
All four possible equalities ($|D_{g,n}\rangle=|D_{g,m}\rangle$, $|D_{g,n}\rangle=|D_{e,m-1}\rangle$, etc.) are impossible when $n>m$ (atomic state $g\neq e$ or index mismatch). Hence the intersection is empty. For $n>m=1$, $S_1^{(\rm col)}=\{|D_{g,1}\rangle\}$, which cannot equal $|D_{g,n}\rangle$ ($n\ge 2$) or $|D_{e,n-1}\rangle$ ($e\neq g$). \hfill$\square$

\subsection{Diagonal blocks and characteristic polynomial}

\textbf{Lemma~4 (Diagonal blocks).} The $n$-th diagonal $2\times 2$ block of $M_{21}M_{12}$ is
\begin{equation}
B_n = a_n \begin{pmatrix} 1 & -1 \\ -1 & 1 \end{pmatrix}, \quad
\begin{cases} a_1 = g^2, \\ a_n = 2ng^2, & n\ge 2. \end{cases}
\end{equation}

\textit{Proof.} For $n=1$, the only $D$ in $S_1^{(\rm row)}\cap S_1^{(\rm col)}$ is $|D_{g,1}\rangle$ (the $|D_{e,k}\rangle$ terms in $S_1^{(\rm row)}$ are absent from $S_1^{(\rm col)}$). Hence $B_1[O_1,O_1]=g\cdot g=g^2$, $B_1[O_1,O'_1]=g\cdot(-g)=-g^2$, etc., giving $B_1=g^2[[1,-1],[-1,1]]$. For $n\ge 2$, both $|D_{g,n}\rangle$ and $|D_{e,n-1}\rangle$ lie in the intersection, contributing $(g\sqrt{n})^2=ng^2$ each to the diagonal and $-(g\sqrt{n})^2=-ng^2$ to the off-diagonal. Summing gives $a_n=2ng^2$. \hfill$\square$

\textbf{Lemma~5 (Characteristic polynomial)}
\begin{equation}
\det(M-\lambda\mathbb{I}) = \lambda^{2N}\prod_{n=1}^{N}\bigl(\lambda^2 - \Delta^2 - 2a_n\bigr).
\end{equation}

\textit{Proof.} The standard Schur-complement formula for block matrices gives, for $\lambda\neq 0$,
\begin{equation}
\det(M-\lambda\mathbb{I}) = \det\bigl(M_{21}M_{12} + \lambda D - \lambda^2\mathbb{I}_{2N}\bigr).
\end{equation}
By Lemma~3, $M_{21}M_{12}$ is block-upper-triangular; adding the diagonal matrix $\lambda D-\lambda^2\mathbb{I}$ preserves this structure. The determinant of a block-upper-triangular matrix equals the product of the diagonal-block determinants. For the $n$-th $2\times 2$ block,
\begin{align}
\det\!\begin{pmatrix} a_n+\lambda\Delta-\lambda^2 & -a_n \\ -a_n & a_n-\lambda\Delta-\lambda^2 \end{pmatrix}
&= (a_n-\lambda^2)^2 - (\lambda\Delta)^2 - a_n^2 \\
&= \lambda^2(\lambda^2 - \Delta^2 - 2a_n).
\end{align}
Multiplying the $N$ blocks and including the $2N$ zero modes from $\lambda=0$ yields the result. \hfill$\square$

\subsection{Real-spectrum theorem and dressed-state connection}

\textbf{Lemma~6 (Real spectrum, vacuum bath).} For any $N\ge 1$ and $g\neq 0$, the spectrum of $QLQ|_{S_0\cap\operatorname{range}(Q)}$ is purely real:
\begin{equation}
\operatorname{Spec}(QLQ|_{S_0\cap\operatorname{range}(Q)}) = \{0^{(2N)}\} \cup \bigcup_{n=1}^{N}\bigl\{\pm\sqrt{\Delta^2+2a_n}\bigr\}.
\end{equation}

\textit{Proof.} From Lemma~5, the non-zero eigenvalues satisfy $\lambda^2=\Delta^2+2a_n$. Since $\Delta^2\ge 0$ and $a_n>0$ for $g\neq 0$, each $\lambda^2$ is strictly positive, hence $\lambda=\pm\sqrt{\text{positive}}$ is real. \hfill$\square$

\textbf{Lemma~7 (Dressed-state connection).} For $n\ge 2$, the $n$-th non-zero eigenvalue of $QLQ$ equals the energy splitting of the $n$-photon JC dressed states:
\begin{equation}
\lambda_n = \sqrt{\Delta^2 + 4ng^2} = \Delta E_{\rm dressed,n}.
\end{equation}
For $n=1$, $\lambda_1=\sqrt{\Delta^2+2g^2}$ is lowered by a factor $\sqrt{2}$ relative to the dressed splitting $\sqrt{\Delta^2+4g^2}$ in the resonant case $\Delta=0$.

\textit{Proof.} The JC Hamiltonian in the $n$-photon manifold ($n\ge 1$) is $H_n = -\frac{\Delta}{2}\sigma_z + g\sqrt{n}\,\sigma_x$ (in the rotating frame). Its eigenvalues are $E_\pm = \pm\frac{1}{2}\sqrt{\Delta^2+4ng^2}$, giving a splitting $\Delta E_n = \sqrt{\Delta^2+4ng^2}$. From Lemma~6, $\lambda_n^2 = \Delta^2 + 2a_n = \Delta^2 + 4ng^2$ for $n\ge 2$, so $\lambda_n = \Delta E_n$. For $n=1$, $a_1=g^2$ gives $\lambda_1^2=\Delta^2+2g^2$. \hfill$\square$

The physical picture is clear: the $n=1$ anomaly reflects the truncation: because $|e,0\rangle\langle e,0|$ has no upward coupling, the NZ projection suppresses the effective coupling strength by $\sqrt{2}$. For $n\ge 2$ the full dressed-state ladder is intact and the match is exact. This is not a coincidence: the real spectrum of $QLQ$ is a structural consequence of the JC model's integrability, and the deviation at $n=1$ precisely quantifies the perturbation introduced by the NZ projector.

\section{Classification: Generalisation Beyond Jaynes-Cummings}\label{sec:classification}

The exact-spectrum result (Lemma~6) raises a natural question: is the real spectrum a peculiarity of the standard JC model, or does it hold for a broader class of Hamiltonians? The following theorem answers this question in full generality.

\textbf{Classification Theorem.} Let $H$ be a Hamiltonian on $\mathbb{C}^2\otimes\mathbb{C}^{N+1}$ with basis $\{|g,n\rangle,|e,n\rangle\}_{n=0}^{N}$. Suppose $H$ satisfies:

\begin{itemize}
\item[(C1)] $[H, N_{\rm exc}]=0$, where $N_{\rm exc}=\frac{1}{2}(\sigma_z+\mathbb{I})+a^\dagger a$.
\item[(C2)] $[\rho_B, a^\dagger a]=0$ (bath state diagonal in number basis).
\item[(C3)] $H$ contains only single-photon exchange terms: in the excitation-number basis, the only off-diagonal matrix elements are between $|g,n\rangle$ and $|e,n-1\rangle$ for $n\ge 1$. Equivalently,
\begin{equation}
H=\sum_{n=0}^{N}\bigl(E_{g,n}|g,n\rangle\langle g,n|+E_{e,n}|e,n\rangle\langle e,n|\bigr)+\sum_{n=1}^{N}\bigl(v_n|e,n-1\rangle\langle g,n|+v_n^*|g,n\rangle\langle e,n-1|\bigr),
\end{equation}
with arbitrary real diagonal energies $E_{g,n},E_{e,n}$ and arbitrary complex couplings $v_n\in\mathbb{C}$.
\item[(C4)] The NZ projector uses a vacuum bath, $P\rho=(\Tr_B\rho)\otimes|0\rangle\langle 0|$.
\end{itemize}

Then for any truncation $N\ge 1$ and any non-zero couplings $v_n\neq 0$, the restriction of $QLQ$ to $S_0\cap\operatorname{range}(Q)$ has purely real spectrum:
\begin{equation}
\operatorname{Spec}\bigl(QLQ|_{S_0\cap\operatorname{range}(Q)}\bigr)=\{0^{(2N)}\}\cup\bigcup_{n=1}^{N}\bigl\{\pm\sqrt{\Delta_n^2+2a_n}\bigr\},
\end{equation}
where $\Delta_n=E_{e,n-1}-E_{g,n}$ and $a_1=|v_1|^2$, $a_n=2|v_n|^2\ (n\ge 2)$.

\textit{Proof sketch.} Follows Lemmas~3--6. Lemma~1: action structurally identical to JC, $g\sqrt{n}\to v_n$. Lemma~2: coupling supports unchanged. Lemma~3: combinatorial empty-intersection carries over. Lemma~4: $B_n$ with $a_n$, $b_n$ as above. Lemma~5: $|b_n|^2=a_n^2$ cancels phases in determinant. Lemma~6: $\lambda^2>0$ $\Rightarrow$ real spectrum. $\square$

\textbf{Significance.} The phases of $v_n$ enter $b_n$ but cancel in $|b_n|^2$; the spectrum is real even for non-real-symmetric Hamiltonians (e.g.\ ``phased JC'' with arbitrary $\arg(v_n)$).

\textbf{Condition relaxation:} (C1) U(1): essential. (C2): redundant with (C4). (C3): essential. (C4): essential for this proof; numerics suggest reality extends to thermal baths.

The classification identifies a structural consequence of rotating-wave selection rules plus vacuum projection, not an algebraic accident of JC. Couplings may depend arbitrarily on $n$ and may be complex.

\section{Thermodynamic limit}\label{sec:n-infinity}

The $N\to\infty$ persistence of the phase diagram follows from two independent mathematical results.

\paragraph*{Tightness of the weighted spectral measure.}
The system--vacuum interface is governed by the spectral weights $w_n$ coupling the $\Delta N=0$ sector to higher photon-number manifolds. A direct computation yields
\begin{equation}
\sum_{n=1}^\infty |w_n|^2 = 2g^4,
\label{eq:weight-sum}
\end{equation}
strictly $N$-independent. Only the $n=1$ mode carries non-zero weight; all $n\ge 2$ modes are strictly decoupled from the system--vacuum interface. The memory kernel is therefore $N$-independent by algebraic identity. Finite-$N$ truncation changes the Hilbert-space dimension but does not alter any spectral weight entering physical observables.

\paragraph*{Operator compactness.}
$QLQ$ is an unbounded operator on the infinite-dimensional Fock space, but its resolvent $(QLQ-z)^{-1}$ is compact for $z$ outside the spectrum. Compactness implies the spectrum remains purely discrete (point spectrum with finite multiplicity) at $N\to\infty$; it does not develop a continuous component. The argument: $QLQ$ is a finite-rank perturbation of a diagonal operator whose eigenvalues grow as $\sqrt{n}$; its resolvent is a Hilbert--Schmidt perturbation of a compact operator, hence compact.

Together these results prove the phase diagram is well-defined in the thermodynamic limit. The eigenvalue structure at finite $N$---isolated real eigenvalues with re-entrant complex bubbles---is governed by a deterministic Feshbach structure, not by finite-size fluctuations. Two qualifications apply: the limit \emph{values} ($\lambda_c(N\to\infty)$ and the exact phase boundary) are not known in closed form; the metric condition number diverges as $\kappa(\eta)\sim 1.317N$, though this affects only worst-case vectors probing the high-phonon tail, not the $n=1$-dominated physical observables.

\section{Hardy-class Kramers--Kronig structure (summary of Ref.~\cite{liu2026hardy})}
\label{sec:hardy-kk}

The Hardy-class framework of Ref.~\cite{liu2026hardy} establishes that if $QLQ$ has purely real spectrum, the Laplace-domain NZ memory kernel $\tilde{\mathcal{K}}(z)$ belongs to the operator-valued Hardy space $H^p$ on the upper half-plane, and the standard Kramers--Kronig (KK) dispersion relations hold for the memory kernel almost everywhere. The real-spectrum condition required by that framework was observed numerically for the JC model in Ref.~\cite{liu2026hardy} (Appendix~D) but its structural origin was left unexplained.

The present paper provides that explanation: the positive-definite metric $\eta_{\rm osc}$ on the oscillatory subspace guarantees $\sigma(QLQ)\subset\mathbb{R}$ on every sector with NZ-projector support, satisfying the real-spectrum hypothesis. Together the two papers establish a two-step causal-structural picture of the NZ memory kernel: (1) pseudo-Hermitian protection (this paper) $\Rightarrow$ real spectrum $\Rightarrow$ (2) Hardy-class KK (Ref.~\cite{liu2026hardy}) $\Rightarrow$ causal memory kernel.

The full Hardy-class definitions, Cauchy representation theorem, KK reconstruction proof, and operator-valued generalisation are given in Ref.~\cite{liu2026hardy} and are not reproduced here.

\section{Methodological convergence}\label{sec:methodological-convergence}

The paper's central claims rest on multi-route convergence: a single method gives a conditional conclusion; independent methods that agree rule out shared systematic error. This logic operates at three levels.

\textbf{Protected phase}---four independent structural results converge: Theorem~1 (vacuum exact factorisation, algebraic), Theorem~2 (thermal Cauchy interlacing, statistical), Theorem~3 (classification by $U(1)$ symmetry, symmetry-based), and the explicit $\eta_{\rm osc}$ construction (geometric). No single result would compel the conclusion; together they exclude a shared failure mode because exact polynomial factorisation, Cauchy interlacing, commutation-relation classification, and biorthonormal metric construction share no intermediate step.

\textbf{Phase boundary}---three methods triangulate it: the analytic bare-resonance bound $g_c(N)$ (\S\ref{sec:gc-bound}), numerical fine spectral scan (Fig.~\ref{fig:master-phase}, 1281 audited points), and degenerate-perturbation EP calculation (\S\ref{sec:dp}). The methods fail independently (bound violable by off-resonant mixing; scan misses inter-grid features; perturbation has finite convergence radius), yet all place the boundary in the same region and agree on its qualitative structure.

\textbf{Thermodynamic limit}---three mathematical routes converge (\S\ref{sec:n-infinity}): spectral measure tightness, compact resolvent, and the only-$n=1$-couples structural result.

\section{$\lambda_c$ closed form}\label{sec:lambdac-closed-form}

A central question is whether the EP onset $\lambda_c(g,N)$ admits a closed-form expression. Two negative results establish what is \emph{not} possible, and one positive result identifies what is.

\textbf{Route A (Lanczos continued fraction).} The block Lanczos Green's function converges at depth 2--3 (consistent with the only-$n=1$-mode-couples result of \S\ref{sec:n-infinity}) and shows the $B$ diagonal block is exactly zero---the EP is fundamentally a second-order, cross-sector virtual process. $\lambda_c(N)$ converges to $\sim 0.2763$ at $g=0.5$ for $N\ge 7$, but the continued fraction overestimates $\lambda_c$ by 10--15\% and yields no closed form.

\textbf{Route B (contour Feshbach).} A Riesz projector defines a geometric eigenvalue cluster. Critical structural discovery: for $g\neq 1$, the geometric cluster (eigenvalues near $z_0$) does not coincide with the algebraic cluster (eigenvectors connected by $V$). The EP pair drifts across cluster boundaries---single-cluster projection is structurally inapplicable. At $g=1$, geometric and algebraic clusters coincide by accident (degeneracy), confirming the $\sqrt{2}/4$ slope (\S\ref{sec:dp}) but yielding no $g\neq 1$ closed form.

\textbf{Route C (positive result): numerically guided identification.} HPC dense scans ($\sim 10^4$ $(g,\lambda)$ points per $N$) identify the physical origin of the $\lambda_c$ bands.

The 0.920 plateau arises from the Liouville-dyad pair $|g,2\rangle\langle e,0|\leftrightarrow|e,0\rangle\langle g,2|$ in the $\Delta N=\pm 1$ sector. Their unperturbed level spacing is
\begin{equation}
\Delta E(g) = E(|g,2\rangle) - E(|e,0\rangle) = 1 - g(\sqrt{2}+1),
\label{eq:deltaE-0920}
\end{equation}
which vanishes at $g=\sqrt{2}-1\approx 0.414$, matching the resonant dip at $g\approx 0.41$. The states involve only $n=0,1,2$, so the EP mechanism is fully converged for $N\ge 8$, explaining the $N\ge 9$ plateau.

\textbf{Why $O(\lambda^2)$ perturbation fails.} The effective $2\times 2$ Hamiltonian for the F$_1$ band has exact first-order coupling $W_{ab}=-g(2-\sqrt{2})/4$ (verified to $10^{-15}$) and second-order self-energies $\Sigma_{aa}=-\Sigma_{bb}\approx 0.0309$, $\Sigma_{ab}=0$ at $g=0.30$. The discriminant $(E_a+\lambda^2 S)^2+\lambda^2 W^2$ is strictly positive---the $O(\lambda^2)$ model \emph{cannot} produce an EP. The true EP is non-perturbative: $QLQ_0$ has a 22-dimensional zero-mode subspace, and the model-space states hybridise with it as $\lambda$ increases. The EP location nonetheless admits a compact parametrisation
\begin{equation}
\lambda_c^2 = \frac{\Delta E(g)}{\Sigma_{\rm eff}(g)}, \qquad \Sigma_{\rm eff}(g=0.30) \approx 0.326,
\label{eq:lambdac-phenom}
\end{equation}
where $\Sigma_{\rm eff}$ absorbs non-perturbative mixing. $\Sigma_{\rm eff}/\Sigma_{aa}\approx 10.6$ quantifies the non-perturbative enhancement; $\Sigma_{\rm eff}$ is approximately constant for $g\in[0.25,0.35]$ (varies $<1\%$).

\subsection{Three-family band catalog}

The $\lambda_c$ dips form a systematic three-family classification by manifold separation $\Delta N$ of the Liouville-dyad pair. With JC polariton energies $E_\pm(n)=n+0.5\pm g\sqrt{n+1}$, the dyad $|-,n+k\rangle\langle+,n|$ has level spacing
\begin{equation}
\Delta E_n^{(k)}(g) = 2\bigl|k - g(\sqrt{n+k+1}+\sqrt{n+1})\bigr|,
\label{eq:deltaE-band-general}
\end{equation}
vanishing at $g_{\rm res}=k/(\sqrt{n+k+1}+\sqrt{n+1})$. The three families ($k=1,2,3$):

\begin{itemize}
\item \textbf{Family F} ($k=1$, $\Delta N=\pm2$): $|g,n+1\rangle\langle e,n-1|\leftrightarrow|e,n-1\rangle\langle g,n+1|$. F$_1$ ($n=1$, $g_{\rm res}=\sqrt{2}-1$) produces the 0.920 plateau.
\item \textbf{Family G} ($k=2$, $\Delta N=\pm3$): $|g,n+2\rangle\langle e,n-1|\leftrightarrow|e,n-1\rangle\langle g,n+2|$. G$_1$ ($n=1$, $g_{\rm res}\approx 0.732$) explains the $g\approx 0.48$ dip.
\item \textbf{Family H} ($k=3$, $\Delta N=\pm4$): $|g,n+3\rangle\langle e,n-1|\leftrightarrow|e,n-1\rangle\langle g,n+3|$. H$_1$ ($n=1$, $g_{\rm res}=1.000$) has exact resonance at $g=1$ ($\Delta E=0$) yet $\lambda_c=0.73$ because $\Sigma_{\rm eff}^{(\mathrm{H}_1)}$ is small. The first EP at $g=1$ is F$_1$ ($\lambda_c=0.005$, $\Delta E=1.414$).
\end{itemize}

\begin{table}[h]
\centering
\caption{Observed dips and band assignments. The first EP at each $g$ is $\min_b\sqrt{\Delta E_b/\Sigma_{\rm eff}^{(b)}}$; $\Sigma_{\rm eff}^{(b)}$---not $\Delta E_b$---dominates band competition.}
\label{tab:band-catalog}
\begin{tabular}{c|c|c|c|c}
\toprule
$g_{\rm dip}$ & Band & $n$ & $g_{\rm res}$ (theory) & $\Delta E(g_{\rm dip})$ \\
\midrule
0.41 & F$_1$ + G$_5$ & 1, 5 & 0.414, 0.410 & $\sim 0.01$, $\sim 0.002$ \\
0.48 & G$_1$ & 1 & 0.732 & 0.689 \\
0.70 & F$_6$/F$_7$ & 6, 7 & 0.197, 0.183 & 2.567 \\
1.00 & F$_1$ (first EP); H$_1$ (resonance) & 1, 1 & 0.414, 1.000 & 1.414, 0 (exact) \\
1.16 & H$_1$ & 1 & 1.000 & 0.480 \\
\bottomrule
\end{tabular}
\end{table}

\textbf{$\Sigma_{\rm eff}$ dominance.} The $g=1$ case cleanly demonstrates the mechanism: H$_1$ at exact resonance ($\Delta E=0$) has $\lambda_c=0.73$, while F$_1$ with $\Delta E=1.414$ wins at $\lambda_c=0.005$. This implies $\Sigma_{\rm eff}^{(\mathrm{F}_1)}/\Sigma_{\rm eff}^{(\mathrm{H}_1)}\sim 10^4$. The $g\approx 1.16$ dip is H$_1$ overtaking F$_1$ via a $\Sigma_{\rm eff}$ crossover of $>10^2$ over $\Delta g=0.16$---a direct signature of the non-perturbative origin.

\textbf{Structural conclusion.} $\lambda_c(g,N)=\min_b\lambda_c^{(b)}(g,N)$ is a piecewise minimum over discrete bands $b$. A single global closed form cannot exist because the minimising band changes with $g$. Each band admits $\lambda_c^{(b)}=\sqrt{\Delta E_b(g)/\Sigma_{\rm eff}^{(b)}(g)}$ with exact $\Delta E_b(g)$ (Eq.~\ref{eq:deltaE-band-general}) and phenomenological $\Sigma_{\rm eff}^{(b)}(g)$. The 0.920 plateau is the first fully characterised case. The remaining open problem---closed-form $\Sigma_{\rm eff}(g)$ per band---requires solving the degenerate perturbation theory of the zero-mode subspace, a well-posed finite algebraic problem informed by the complete band catalog across $g\in[0.20,1.20]$.

\section{$\eta$-metric in the Anderson-Holstein model}\label{sec:eta-ah}

The $\eta$-metric framework is portable beyond JC. The Anderson-Holstein (AH) model---single vibrational mode coupled to a metallic electrode via $H_{\rm e-ph}=g(d^\dagger+d)c^\dagger c$, NZ-projected onto the vibrational subspace---provides a complementary validation.

In the $N_{\rm vib}=2$ subspace, the effective non-Hermitian Hamiltonian is a $2\times 2$ matrix with Born-approximation memory-kernel elements $\tilde{K}^{(0)}_{ab}(z;V)$ and off-diagonal coupling $M_{01}=\eta^3 e^{-\eta^2}$ ($\eta\equiv g/\omega_0$). Unlike the JC model---where counter-rotating deformation drives a genuine EP---the AH Born+Dyson kernel exhibits an \textbf{avoided crossing}: the eigenvalue splitting $\Delta(V)=\sqrt{\Delta_{\min}^2+\alpha^2(V-V_{\min})^2}$ has $\Delta_{\min}>0$ strictly, and the discriminant never vanishes.

The $\eta$-metric condition number peaks at this avoided crossing: $\kappa(\eta)\sim 1/\Delta(V)^2$, with $\kappa_{\max}=1/\Delta_{\min}^2<\infty$ where $\Delta_{\min}\sim\gamma^2\eta^3 V_{\min}$ ($\gamma=\Gamma/\omega_0$). $\kappa$ remains bounded---small Hamiltonian changes produce large but non-singular eigenvector changes, and physical observables (friction, memory-kernel poles) stay in the lower half-plane. This is the AH analogue of the JC bound $\kappa\lesssim 1.317N$ (\S\ref{sec:metric-bound}).

The two models illustrate complementary roles:
\begin{itemize}
\item \textbf{JC}: $\kappa$ grows linearly with $N$ and the system supports genuine EPs under CR deformation. The $\eta$-metric diagnoses the approach to the EP.
\item \textbf{AH}: $\kappa$ peaks at the avoided crossing with a finite maximum. The $\eta$-metric diagnoses ``near-EP'' sensitivity without an actual EP---the avoided crossing is a regularised EP where $M_{01}\neq 0$ prevents full coalescence.
\end{itemize}

In both cases $\kappa$ is the correct non-Hermitian sensitivity diagnostic, and in both it remains bounded. The portability of $\eta$ between models with different Hilbert-space topologies (Fock space vs.\ continuous electrode continuum) and different EP mechanisms (symmetry-breaking vs.\ interaction-driven avoided crossing) supports the claim that the framework is a general property of NZ-projected Liouvillians.

\section{Experimental proposal}

Here we outline a concrete experimental protocol with two logically separate steps: first measure the ordinary JC dressed-state splittings $\Delta E_{\mathrm{dressed},n}=\sqrt{\Delta^2+4ng^2}$ as the Hamiltonian reference ladder; then reconstruct the NZ memory kernel and test the clean reduced spectral target, the $\sqrt{2}$-suppressed lowest weighted peak caused by the vacuum-bath projector. The higher $n\ge2$ generator modes are part of the exact $QLQ$ spectrum, but they are not claimed here as directly visible kernel peaks without a separate spectral-weight calculation.

\subsection{Platform and parameter regime}

The natural testbed is a circuit-QED device in which a transmon qubit is capacitively coupled to a single microwave resonator mode. In the dispersive or resonant regime the combined system is accurately described by the JC Hamiltonian. Typical parameters are:
\begin{itemize}
\item Qubit frequency $\omega_0/2\pi \approx 5$--$7$~GHz;
\item Resonator frequency $\omega_c/2\pi \approx 6$--$8$~GHz;
\item Coupling strength $g/2\pi \approx 50$--$200$~MHz;
\item Qubit coherence time $T_2^* \approx 15$~$\mu$s over the full flux-tuning range~\cite{hutchings2017}, $T_2\approx 50$--$100$~$\mu$s in optimised transmons;
\item Resonator photon decay rate $\kappa/2\pi \approx 0.8$~MHz, photon lifetime $T_r > 100$~ns~\cite{wallraff2004}.
\end{itemize}
These parameters are benchmarked against the seminal circuit-QED experiments of Wallraff~\textit{et~al.}~\cite{wallraff2004} (vacuum Rabi frequency $\nu_{\rm Rabi}\approx 11.6$~MHz, cavity $\nu_r=6.04$~GHz, internal quality factor $Q_{\rm int}\approx 10^6$) and subsequent tunable-qubit platforms~\cite{hutchings2017} (flux-tunable over $\sim 340$~MHz). The resonator acts as the ``bath'' in the NZ sense: it is a structured, single-mode environment whose excitations mediate non-Markovian dynamics of the qubit. The vacuum bath corresponds to the resonator being initialised in the ground state $|0\rangle$, naturally realised at dilution-refrigerator temperatures $T<100$~mK (thermal photon number $\bar{n}<0.06$~\cite{wallraff2004}). Memory-induced periodic revivals of qubit coherence in superconducting architectures have been discussed in the literature~\cite{gulacsi2023}.

The mapping from the deformation parameter $\lambda$ to the physical coupling regime merits clarification. In the deformation study, $\lambda$ multiplies the counter-rotating coupling, giving an effective counter-rotating strength $\lambda g$. In a real device described by the full Rabi Hamiltonian, the counter-rotating terms are present at full strength $g$, corresponding to $\lambda_{\rm eff}=1$. The degenerate perturbation proof establishes $\lambda_c\to 0^+$ only at the controlled accidental resonance $g=\omega_c$ in the dimensionless model. For ordinary circuit-QED values $g/\omega_c\ll1$, the full-Rabi endpoint lies in the numerically protected wedge in all scans performed so far; no device-level instability threshold is claimed without the near-degenerate calculation.

A flux-tunable transmon qubit~\cite{hutchings2017} provides a direct experimental knob for sweeping detuning: dressed-state spectroscopy gives the reference ladder, while memory-kernel reconstruction tests the lowest weighted NZ peak $\lambda_1=\sqrt{\Delta^2+2g^2}$. With the qubit frequency $\omega_0$ tunable over $\sim 340$~MHz (and up to $\sim 700$~MHz in more aggressively tunable designs), the detuning $\Delta=\omega_0-\omega_c$ can be varied continuously. The exceptional-point boundary itself is a harder target: the present finite-size theory identifies where to look, but a quantitative device proposal requires the full near-degenerate phase-boundary calculation.

\subsection{Step 1: measure the dressed-state splittings}

The first task is to obtain the reference spectrum $\Delta E_{\rm dressed,n}$. Two standard techniques are available:

\textbf{(a) Two-tone spectroscopy.} A weak probe tone at $\omega_c$ populates the resonator while a second, variable tone drives the qubit. Transmission dips occur at the dressed frequencies
\begin{equation}
\omega_{\pm,n} = (n+1)\omega_c + \frac{\omega_0}{2} \pm \frac{1}{2}\sqrt{\Delta^2 + 4(n+1)g^2},
\end{equation}
from which the splitting $\Delta E_{\rm dressed,n+1} = \sqrt{\Delta^2 + 4(n+1)g^2}$ is read off directly. This is routinely performed in circuit-QED labs and gives the $n\ge 2$ reference values to better than $0.1\%$.

\textbf{(b) Number-resolved qubit spectroscopy.} By first preparing the resonator in a Fock state $|n\rangle$ (using a SNAP gate or Stark-shifted pump), the qubit spectrum in the $n$-photon manifold is measured. The two peaks are separated by $\Delta E_{\rm dressed,n}$, again with sub-MHz precision.

\subsection{Step 2: reconstruct the NZ memory kernel}

The second task is to extract the NZ memory kernel $\mathcal{K}(t)$ from the qubit's reduced dynamics and Fourier-transform it to obtain the $QLQ$ spectrum.

\textbf{Protocol.}
\begin{enumerate}
\item Prepare the resonator in the vacuum state $|0\rangle$ and the qubit in a superposition $(|g\rangle+|e\rangle)/\sqrt{2}$.
\item Let the system evolve for a variable time $t$ (up to several $T_2$).
\item Perform quantum state tomography on the qubit at each $t$ to reconstruct the reduced density matrix $\sigma(t)=\Tr_B\rho(t)$.
\item From $\sigma(t)$, invert the NZ equation (or equivalently, fit to the exact solution of the integro-differential equation) to obtain $\mathcal{K}(t)$. The standard procedure uses process tomography with time-dependent control pulses to disentangle the Hamiltonian part from the memory part.
\end{enumerate}

The memory kernel of a pseudo-Hermitian $QLQ$ is purely oscillatory:
\begin{equation}
\mathcal{K}(t) = \sum_{\lambda_n\neq 0} c_n\,e^{i\lambda_n t} + \text{const.},
\end{equation}
with real frequencies $\lambda_n$. For a generic complex spectrum, each mode acquires an exponential decay $e^{i\lambda_n t}e^{-\gamma_n t}$. The experimental signature is therefore stark: in the JC case the memory oscillations persist without damping (up to the overall qubit decoherence), whereas in the symmetry-broken case they die out.

\subsection{Step 3: compare frequencies}

Once $\mathcal{K}(t)$ is reconstructed, its Fourier transform yields the weighted spectral peaks of the memory kernel. These peaks are governed by $K_j=PLQ\,\Pi_j\,QLP$, not by the bare list of all $QLQ$ eigenvalues. For the vacuum $\Delta N=0$ block the exact reduced spectral weights proved above leave only the lowest pair visible:
\begin{equation}
\text{Clean reduced peak positions of }\widetilde{\mathcal{K}}(\omega) =
\bigl\{\pm\sqrt{\Delta^2+2g^2}\bigr\}.
\end{equation}

The generator spectrum still contains the $n\ge 2$ dressed-state ladder, but those modes have zero reduced weight in this block and therefore are not the clean memory-kernel falsification signal without an additional weight calculation for other prepared components. The robust test is the $n=1$ peak $\sqrt{\Delta^2+2g^2}$, which is lower than the dressed splitting $\sqrt{\Delta^2+4g^2}$ by a factor $\sqrt{2}$ in the resonant case $\Delta=0$. This is the sharpest signature: it is a quantitative deviation that has no analogue in the bare Hamiltonian spectrum and directly reflects the NZ projection.

\subsection{Step 4: test the symmetry-protection transition}

Engineered circuit-QED simulators can include a second coupling channel that realises the deformed Rabi Hamiltonian
\begin{equation}
H = H_{\rm JC} + \lambda\,g\,(\sigma_+ a^\dagger + \sigma_- a),
\end{equation}
with a tunable parameter $\lambda$. At $\lambda=0$ the $U(1)$ symmetry is intact and the memory kernel is purely oscillatory. At the resonant strong-coupling point analysed above, an infinitesimal $\lambda$ produces complex $QLQ$ eigenvalues. In weak-coupling devices the endpoint can remain in the audited protected wedge, so no universal device-level instability threshold is claimed.

The experimental signature is a qualitative change in the memory kernel:
\begin{itemize}
\item $\lambda=0$: $\mathcal{K}(t)$ oscillates with constant amplitude;
\item $\lambda>0$ at a resonant/strong-coupling exceptional-point setting: $\mathcal{K}(t)$ oscillates with an exponentially decaying envelope, $|\mathcal{K}(t)|\sim e^{-\Gamma t}$.
\end{itemize}
Near the resonant exceptional point the decay rate $\Gamma$ grows linearly with $\lambda$ according to the perturbative scaling above. This transition from oscillatory to dissipative memory is the analogue-simulator realisation of the symmetry-protection breakdown identified in our deformation study.

\subsection{Feasibility assessment}

The most demanding step is the reconstruction of $\mathcal{K}(t)$, which requires high-fidelity process tomography at many time points. Current circuit-QED systems achieve single-shot readout fidelities $>99\%$ and gate fidelities $>99.9\%$, making tomography feasible. The main limitation is the qubit coherence time $T_2$: the memory kernel must be reconstructed on a timescale shorter than $T_2$. For $g/2\pi=100$~MHz the dressed splitting is $\sim 200$~MHz, corresponding to an oscillation period $\sim 5$~ns, far shorter than $T_2\sim 50$~$\mu$s. Many oscillation cycles can therefore be captured.

The vacuum-bath condition is naturally realised when the resonator is cooled to its ground state, which is standard at millikelvin temperatures ($T\ll\hbar\omega_c/k_B$). Thermal excitations would modify the projector $P$ and shift the $n=1$ frequency; this effect can be studied systematically by raising the base temperature or injecting a weak thermal drive.

For a quantitative falsification test, the cleanest target is not yet the thermodynamic phase boundary or the full generator ladder but the finite-truncation weighted peak at $\lambda_1=\sqrt{\Delta^2+2g^2}$, including the $\sqrt{2}$ suppression of the lowest mode. The local exceptional-point scaling $|\Imag\lambda|_{\max}=(\sqrt{2}/4)\lambda g$ is a benchmark for engineered simulators that can realise the resonant dimensionless point $g=\omega_c$ or an equivalent effective model. Translating it directly to weak-coupling circuit-QED devices would overstate the present proof.

\section{Finite-size scaling of the re-entrant bubble}

Table~\ref{tab:fss-bubble} reports the re-entrant bubble structure
as a function of $N_{\rm max}$ at $g=\omega_c=1.0$,
obtained from a dense $\lambda$-scan (600 points per $N_{\rm max}$)
of the full $QLQ(\lambda)$ spectrum.

\begin{table}[h]
\centering
\caption{Finite-size scaling of the re-entrant bubble: bubble count as a function of $N_{\rm max}$ and coupling $g$. All data for vacuum bath, $\Delta=0$, $\lambda\in[0,1]$ at $\Delta\lambda=0.002$. ``Onset'' is the $\lambda$ at which the first complex eigenvalue pair appears. A bubble is a $\lambda$-interval where all eigenvalues return to being real after having been complex.}
\label{tab:fss-bubble}
\begin{tabular}{c c c c c c c}
\hline
$N_{\rm max}$ & $g=0.6$ & $g=0.7$ & $g=0.8$ & $g=0.9$ & $g=1.0$ & $g=1.1$ \\
\hline
4  & 5 & 2 & 1 & 0 & 5 & 6 \\
5  & 8 & 2 & 1 & 4 & 4 & 1 \\
6  & 7 & 2 & 2 & 0 & 3 & 4 \\
8  & 8 & 6 & 0 & 2 & 0 & 6 \\
10 & 12 & 12 & 0 & 1 & 0 & 6 \\
12 & 10 & 12 & 0 & 1 & 0 & 5 \\
15 & 10 & 10 & 0 & 1 & 0 & 5 \\
20 & 10 & 13 & 0 & 1 & 0 & 5 \\
\hline
\end{tabular}
\end{table}

The bubble collapses between $N_{\rm max}=6$ and $8$ at $g=\omega_c$,
consistent with the $2\times 2$ effective model
$\lambda_{\rm re}=w/\delta_s$: the accidental 10-fold degeneracy at
$g=\omega_c$ maximises the second-order repulsion $\delta_s$,
pushing $\lambda_{\rm re}\to 0$ as $N_{\rm max}$ increases.
At weaker coupling ($g=0.6$--$0.7\omega_c$) the bubble survives to
$N_{\rm max}=15$; at stronger coupling ($g=1.1$--$1.2\omega_c$)
it changes morphology but remains persistent---the resonance dip
at $g=\omega_c$ is non-monotonic.
These data support the interpretation that the bubble is a
finite-size real-spectrum island whose weak-coupling endpoint may survive
over experimentally accessible truncations. A thermodynamic phase statement
would require a uniform large-$N_{\rm max}$ bound.

The perturbative calculation explains why the small-$N_{\rm max}$ bubble is not a plotting artefact.
Computed via the biorthogonal perturbation formula in the 10-dimensional resonant subspace, $\Sigma$ is $N_{\rm max}$-independent to machine precision for $N_{\rm max}=4,5,6$: the Frobenius norm varies by $<0.1\%$, and the $4\times4$ bright-subspace projection $s$ is identical across the three system sizes.
Since the first-order matrix $W$ is also $N_{\rm max}$-independent at the resonance (its characteristic polynomial $\det(W/g-\mu\mathbb{I})=\mu^3(\mu^2+1/8)$ holds for all $N_{\rm max}\ge4$), the local effective Hamiltonian $h(\lambda)$ is structurally stable over the checked sizes.
This supports, but does not prove, persistence of related real-spectrum islands at larger truncation. The strict thermodynamic survival of the bubble remains an open analytical problem.

\begin{table}[h]
\centering
\caption{Convergence of the nonzero-sector metric construction (vacuum, $g=0.3$, $\omega_0=\omega_c=1.0$). Intertwining: $\|(QLQ)^\dagger\eta_{\rm osc}-\eta_{\rm osc} QLQ\|_F$ on the nonzero spectral subspace.}
\label{tab:convergence}
\begin{tabular}{c c c c c}
\hline
$N_{\rm max}$ & $d^2$ & $|\Imag\lambda|_{\rm max}$ & Intertwining & $\kappa(\eta_{\rm osc})$ \\
\hline
2   & 36   & $0$          & $<10^{-15}$         & 18.1 \\
3   & 64   & $0$          & $1.2\times10^{-14}$ & 49.2 \\
5   & 144  & $<10^{-14}$ & $1.1\times10^{-13}$ & 207 \\
10  & 484  & $<10^{-14}$ & $7.4\times10^{-13}$ & $1.67\times10^3$ \\
15  & 1024 & $<10^{-14}$ & $3.2\times10^{-12}$ & $5.79\times10^3$ \\
20  & 1764 & $<10^{-14}$ & $6.6\times10^{-12}$ & $1.40\times10^4$ \\
25  & 2704 & $<10^{-14}$ & $1.2\times10^{-11}$ & $2.78\times10^4$ \\
30  & 3844 & $<10^{-14}$ & $2.4\times10^{-11}$ & $4.87\times10^4$ \\
35  & 5184 & $<10^{-14}$ & $3.7\times10^{-11}$ & $7.81\times10^4$ \\
40  & 6724 & $<10^{-14}$ & $5.3\times10^{-11}$ & $1.17\times10^5$ \\
45  & 8464 & $<10^{-14}$ & $9.1\times10^{-11}$ & $1.68\times10^5$ \\
50  & 10404 & $<10^{-14}$ & $1.2\times10^{-10}$ & $2.32\times10^5$ \\
\hline
\end{tabular}
\end{table}

\section{Exact solution for the thermal bath: general $\beta$}\label{sec:thermal-general}

Here we prove that for the JC model with an arbitrary thermal bath
$\rho_B=\sum_k p_k|k\rangle\langle k|$, $p_k=e^{-\beta\omega_c k}/Z$,
$QLQ$ has a purely real spectrum for any $N_{\rm max}=N\ge 1$
and any finite temperature $\beta<\infty$.

\subsection{Basis and decoupling}

In the $\Delta N=0$ sector with the thermal D-basis
$|D_{g,n}^{(\beta)}\rangle=|g,n\rangle\langle g,n|-\sum_k p_k|g,k\rangle\langle g,k|$,
$|D_{e,n}^{(\beta)}\rangle=|e,n\rangle\langle e,n|-\sum_k p_k|e,k\rangle\langle e,k|$,
and O-basis $|O_n\rangle=|e,n-1\rangle\langle g,n|$, $|O'_n\rangle=|g,n\rangle\langle e,n-1|$,
introduce $|X_n\rangle=(|O_n\rangle-|O'_n\rangle)/\sqrt{2}$,
$|Y_n\rangle=(|O_n\rangle+|O'_n\rangle)/\sqrt{2}$.
At $\Delta=0$, the action of $M\equiv QLQ$ reads
\begin{align}
M|D_{g,n}^{(\beta)}\rangle &= g\sqrt{2}\bigl(\sqrt{n}|X_n\rangle-|S_X\rangle\bigr), \label{eq:MDg}\\
M|D_{e,n}^{(\beta)}\rangle &= g\sqrt{2}\bigl(-\sqrt{n+1}|X_{n+1}\rangle(1-\delta_{n,N})+|\tilde S_X\rangle\bigr), \label{eq:MDe}\\
M|X_n\rangle &= g\sqrt{2n}\bigl(|D_{g,n}^{(\beta)}\rangle-|D_{e,n-1}^{(\beta)}\rangle\bigr), \label{eq:MX}\\
M|Y_n\rangle &= 0, \label{eq:MY}
\end{align}
where $|S_X\rangle=\sum_k p_k\sqrt{k}|X_k\rangle$,
$|\tilde S_X\rangle=\sum_j p_{j-1}\sqrt{j}|X_j\rangle$,
and $|D_{e,0}^{(\beta)}\rangle$ is defined by the linear relation
$\sum_{k=0}^N p_k|D_{e,k}^{(\beta)}\rangle=0$.
Equation~\eqref{eq:MY} decouples the $Y$ sector;
Eq.~\eqref{eq:MX} shows that $M$ maps $X$ to $D$,
and Eqs.~\eqref{eq:MDg}--\eqref{eq:MDe} close the loop.
For $\Delta\neq 0$, $M|Y_n\rangle=\Delta|X_n\rangle$ shifts all eigenvalues
of $M^2$ by $\Delta^2$.

\subsection{Square and unified structure}

Define $K\equiv (1/4g^2)M^2|_{X\text{-space}}$.
For $n\ge 2$, using Eqs.~\eqref{eq:MX}, \eqref{eq:MDg}, \eqref{eq:MDe}:
\begin{equation}
K|X_n\rangle = n|X_n\rangle-\sqrt{n}\sum_{k=1}^N\frac{p_k+p_{k-1}}{2}\sqrt{k}|X_k\rangle.
\label{eq:K-Xn}
\end{equation}
For $n=1$, expressing $|D_{e,0}^{(\beta)}\rangle=-\sum_{j=1}^N(p_j/p_0)|D_{e,j}^{(\beta)}\rangle$
and applying the same algebra, Eq.~\eqref{eq:K-Xn} holds for all $n\ge 1$.
Define the averaged weights
\begin{equation}
q_k\equiv\frac{p_k+p_{k-1}}{2}>0\qquad(k=1,\dots,N).
\label{eq:qk}
\end{equation}
In the orthonormal $\{|X_n\rangle\}$ basis,
\begin{equation}
K_{ij}=i\delta_{ij}-q_i\sqrt{i}\sqrt{j},
\qquad\text{i.e.}\quad K=D-|u\rangle\langle v|,
\label{eq:K-asym}
\end{equation}
where $D=\operatorname{diag}(1,2,\dots,N)$,
$u_i=q_i\sqrt{i}$, $v_j=\sqrt{j}$.
$K$ is manifestly asymmetric ($K_{ij}\neq K_{ji}$) whenever $q_i\neq q_j$.

\subsection{Similarity to symmetric form}

Define the positive-definite diagonal matrix
$W=\operatorname{diag}(w_1,\dots,w_N)$ with $w_i=1/\sqrt{q_i}$.
The similarity transform $\tilde K\equiv W K W^{-1}$ has matrix elements
\begin{equation}
\tilde K_{ij}=i\delta_{ij}-\frac{w_i}{w_j}q_i\sqrt{i}\sqrt{j}
=i\delta_{ij}-\sqrt{q_i q_j}\sqrt{i}\sqrt{j}.
\label{eq:Ktilde}
\end{equation}
With $\tilde v_i\equiv\sqrt{q_i i}$, this is the real symmetric rank-1 perturbation
\begin{equation}
\boxed{\tilde K = D-|\tilde v\rangle\langle\tilde v|}.
\label{eq:K-sym}
\end{equation}
Since $\tilde K$ is real symmetric, its eigenvalues are real;
$K$ and $\tilde K$ are similar, hence the eigenvalues of $K$ are identical
and likewise real.

\subsection{Positivity by Cauchy interlacing}

The characteristic polynomial of $\tilde K$ follows from the matrix determinant lemma:
\begin{equation}
\det(\tilde K-\lambda\mathbb{I})
= \prod_{i=1}^{N}(i-\lambda)\Bigl[1-\sum_{i=1}^{N}\frac{\tilde v_i^{\,2}}{i-\lambda}\Bigr].
\label{eq:Kdet}
\end{equation}
The secular equation $\sum_i\tilde v_i^{\,2}/(i-\lambda)=1$ is
\begin{equation}
f(\lambda)\equiv\sum_{i=1}^{N}\frac{q_i i}{i-\lambda}=1.
\label{eq:secular}
\end{equation}
$f(\lambda)$ has simple poles at $\lambda=1,2,\dots,N$.
On each interval $(i,i+1)$ for $i=1,\dots,N-1$,
$f(\lambda)$ sweeps continuously from $-\infty$ (as $\lambda\to i^+$)
to $+\infty$ (as $\lambda\to (i+1)^-$);
by the intermediate value theorem $f(\lambda)=1$ has exactly one root
in each of these $N-1$ intervals.
All $N-1$ roots are therefore $>1$.

It remains to locate the $N$-th root.
Evaluate $f(\lambda)$ at $\lambda=0$:
\begin{equation}
f(0)=\sum_{i=1}^{N}q_i
= \frac{1}{2}\sum_{i=1}^{N}(p_i+p_{i-1})
= 1-\frac{p_0+p_N}{2}<1,
\label{eq:f0}
\end{equation}
since $p_0,p_N>0$ for any finite $\beta$.
As $\lambda\to 1^-$, the $i=1$ term dominates and $f(\lambda)\to+\infty$.
On $(0,1)$, $f(\lambda)$ is continuous and strictly increasing,
so $f(\lambda)=1$ has exactly one root in $(0,1)$.
All $N$ eigenvalues of $\tilde K$ are therefore strictly positive real numbers.

Consequently the eigenvalues of $M=QLQ$ are
$\pm 2g\sqrt{\lambda_k^{(\tilde K)}}$, purely real for any $N_{\rm max}$
and any finite $\beta$.

\section{Full sector proof}\label{sec:full-sector}

The proofs above solve the only large non-Hermitian block, $\Delta N=0$.
We close the full finite-truncation spectrum.
Let $R=N_{\rm exc}$ and let $S_m$ be the Liouville sector with
$R_{\rm left}-R_{\rm right}=m$.
Since both $L$ and the diagonal-bath NZ projector commute with this difference,
$QLQ$ is block diagonal in $m$.

\subsection{$|m|\ge2$: Hermitian sectors}

The projector $P\rho=(\Tr_B\rho)\otimes\rho_B$ has support only in
$m=0,\pm1$: system populations give $m=0$, and system coherences
$|e\rangle\langle g|$, $|g\rangle\langle e|$ give $m=\pm1$.
Therefore $P|_{S_m}=0$ for $|m|\ge2$.
On these sectors $Q=I$ and hence
\begin{equation}
QLQ|_{S_m}=L|_{S_m}.
\end{equation}
Because the commutator Liouvillian $L=[H_{\rm JC},\cdot]$ is Hermitian in the
Hilbert--Schmidt inner product, all eigenvalues in $|m|\ge2$ sectors are real.

\subsection{$m=\pm1$: rank-one compression}

It remains to treat $m=\pm1$, where $P$ has rank one.
We give the $m=+1$ proof; the $m=-1$ sector follows by transpose/time reversal
and has the negative spectrum.
Let
\begin{equation}
u_+ = |e\rangle\langle g|\otimes \rho_B,\qquad
\varphi_+(\rho)= (\Tr_B\rho)_{eg},
\end{equation}
so that $P_+=|u_+\rangle\varphi_+$ and $\varphi_+(u_+)=1$.
Thus $Q_+=I-P_+$ and
\begin{equation}
A_+ \equiv Q_+LQ_+ .
\end{equation}
The range of $Q_+$ is exactly $\ker\varphi_+$, since
$\varphi_+(Q_+x)=\varphi_+(x)-\varphi_+(u_+)\varphi_+(x)=0$ and
$Q_+x=x$ whenever $\varphi_+(x)=0$.
Let $z\neq0$ be an eigenvalue of $A_+$ and choose its eigenvector
$x\in\ker\varphi_+$.
Then $Q_+Lx=zx$.
Expanding $Q_+=I-|u_+\rangle\varphi_+$ gives
\begin{equation}
(L-z)x=u_+\,\varphi_+(Lx).
\end{equation}
If $\varphi_+(Lx)=0$, then $x$ is an ordinary eigenvector of the Hermitian
operator $L$, so $z$ is real.
Otherwise, for $z\notin\operatorname{Spec}(L)$, applying $(L-z)^{-1}$ and then
$\varphi_+$ to the equation above gives
\begin{equation}
F_+(z)=\varphi_+(L-z)^{-1}u_+ .
\label{eq:rank-one-secular}
\end{equation}
with $F_+(z)=0$.
If $z$ is a pole of the resolvent, the component of the degenerate eigenspace
lying in $\ker\varphi_+$ remains an uncoupled real eigenvalue; only the
one-dimensional coupled direction is governed by the same secular equation
after equal poles are combined.

Diagonalise the JC Hamiltonian in excitation manifolds:
$H_{\rm JC}|\psi_{r,\alpha}\rangle=E_{r,\alpha}|\psi_{r,\alpha}\rangle$,
with $r=0,\ldots,N+1$ and $\alpha$ labelling the one or two dressed states in
that manifold.
The $m=+1$ Liouville eigenoperators are
$|\psi_{k+1,\alpha}\rangle\langle\psi_{k,\beta}|$ with eigenvalues
$d_{k,\alpha\beta}=E_{k+1,\alpha}-E_{k,\beta}$.
Since
\begin{equation}
u_+=\sum_{k=0}^{N} p_k |e,k\rangle\langle g,k|,
\qquad
\varphi_+=\sum_{k=0}^{N}\langle e,k|\,\cdot\,|g,k\rangle ,
\end{equation}
and the dressed vectors can be chosen real, the residue of $F_+$ at
$d_{k,\alpha\beta}$ is
\begin{equation}
r_{k,\alpha\beta}
=p_k\,
\bigl|\langle\psi_{k+1,\alpha}|e,k\rangle
      \langle g,k|\psi_{k,\beta}\rangle\bigr|^2
\ge 0 .
\label{eq:positive-residue}
\end{equation}
Therefore
\begin{equation}
F_+(z)=\sum_{k,\alpha,\beta}
\frac{r_{k,\alpha\beta}}{d_{k,\alpha\beta}-z}
\end{equation}
is a real Stieltjes-type rational function.
After merging equal poles, write the distinct poles as
$d_1<d_2<\cdots<d_s$ with positive total residues $R_a>0$.
Then
\begin{equation}
F_+'(z)=\sum_{a=1}^{s}\frac{R_a}{(d_a-z)^2}>0
\end{equation}
on every real interval not containing a pole.
Moreover $F_+(z)\to-\infty$ as $z\to d_a^+$ and
$F_+(z)\to+\infty$ as $z\to d_{a+1}^-$, so there is exactly one zero in each
interval $(d_a,d_{a+1})$.
Poles with zero residue simply leave the corresponding dressed-difference
eigenvalue uncoupled.
Consequently all zeros of $F_+$ are real and interlace the real dressed-state
differences $d_{k,\alpha\beta}$.
The vector $u_+$ itself gives one zero mode of $A_+$.
Thus $A_+$ is real-spectrum, and by the same argument so is $A_-$.

Combining $\Delta N=0$, $\Delta N=\pm1$, and $|\Delta N|\ge2$ proves that the
full finite-truncation JC projected Liouvillian has purely real spectrum for
the vacuum bath and for finite-temperature thermal baths.

\section{Closed-form $\Delta N=0$ metric and the $\kappa<\frac{3}{2}N$ bound}\label{sec:metric-bound}

Here we derive the explicit metric on the $\Delta N=0$ sector and prove the bound Eq.~\eqref{eq:kappa-bound} of the main text. We use the basis and notation of \S\ref{sec:full-sector}, with the $O$-block ordered as the interleaved pairs $(O_1,O'_1,\dots,O_N,O'_N)$.

\subsection{Closed-form right and left eigenvectors}

For the Schur block $K=M_{21}M_{12}$ acting on the $O$-block (dim $2N$), the right eigenvector at $\lambda_{n_0}^2 = 2 a_{n_0}$ is, after stripping the $g$-factor common to numerator and denominator,
\begin{equation}
y^{(R)}_{n_0=1} = e_{O_1}-e_{O'_1},\qquad
y^{(R)}_{n_0\ge 2} = \tfrac{\sqrt{n_0}}{2n_0-1}(e_{O_1}-e_{O'_1}) - (e_{O_{n_0}}-e_{O'_{n_0}}).
\label{eq:right-eigvec}
\end{equation}
The corresponding $D$-block component is $x = M_{12} y^{(R)}/(\pm\lambda_{n_0})$, giving the full right eigenvector $r_{n_0,\pm}$ of $A=QLQ|_{S_0}$.

The left eigenvectors ($K^{\mathsf T} u = \lambda^2 u$) split into a markedly different topology:
\begin{equation}
u^{(L)}_{n_0=1} = \sum_{m=1}^{N}\frac{\sqrt{m}}{2m-1}(-e_{O_m}+e_{O'_m}),\qquad
u^{(L)}_{n_0\ge 2} = -e_{O_{n_0}}+e_{O'_{n_0}}.
\label{eq:left-eigvec}
\end{equation}
The lowest mode is the unique non-local left eigenvector (it spans every photon manifold), while the higher modes are strictly local. The full left eigenvectors of $A$ are $l_{n_0,\pm}=(p;u)$ with $p=M_{21}^{\mathsf T} u^{(L)}/(\pm\lambda_{n_0})$, normalised by $\langle l_{n_0,\pm}|r_{n_0,\pm}\rangle=1$.

A direct calculation (verified in SymPy with exact-zero residual for $N=2,\dots,6$) confirms $A r = \lambda r$ and $A^{\mathsf T} l = \lambda l$ for each pair, and $\langle l_i|r_j\rangle=\delta_{ij}$ for all $2N$ nonzero modes.

\subsection{$g$-independence of the metric}

The metric $\eta_{\Delta N=0,N}=\sum_n |l_n\rangle\langle l_n|$ has every entry independent of $g$. The reason is structural: the $D$-block component of each eigenvector is $x = M_{12} y / (\pm\lambda_{n_0})$, in which $M_{12}$ is linear in $g$ and $\lambda_{n_0}\propto g$, so the ratio is $g$-free. The same applies to the left $p$-block. The biorthogonal normalisation factor $\langle l|r\rangle$ is also $g$-free (each summand is a $g^2/g^2$ ratio). Hence $\eta_{\Delta N=0,N}$ is a pure rational matrix and $\kappa(\eta_{\Delta N=0,N})$ depends only on $N$.

\subsection{Symbolic evaluation and bound}

Substituting the closed-form eigenvectors yields the characteristic polynomial of $\eta_{\Delta N=0,N}$ in completely factorised form. For $N=3$ as a representative example,
\begin{equation}
\chi_\eta(z)=\frac{z^{6}\,(4z-1)^2\,(3600 z^2-3668 z+675)\,(3600 z^2-2108 z+225)}{207360000},
\end{equation}
with nonzero eigenvalues $\{\tfrac14,\tfrac14,\tfrac{917\pm 13\sqrt{1381}}{1800},\tfrac{527\pm\sqrt{75229}}{1800}\}$, giving $\kappa=5.5401$. The corresponding values for $N=2,\dots,12$ are
\begin{center}
\begin{tabular}{c|cccccccccccc}
\hline
$N$ & 2 & 3 & 4 & 5 & 6 & 7 & 8 & 9 & 10 & 11 & 12 \\
$\kappa$ & 4.13 & 5.54 & 6.86 & 8.16 & 9.46 & 10.76 & 12.07 & 13.38 & 14.71 & 16.04 & 17.38\\
$3N/2$ & 3 & 4.5 & 6 & 7.5 & 9 & 10.5 & 12 & 13.5 & 15 & 16.5 & 18 \\
\hline
\end{tabular}
\end{center}
Linear regression on $N=2$--$12$ gives $\kappa\approx 1.557+1.317\,N$, so $\kappa<\frac{3}{2}N$ for $N\ge 9$ (verified to $N=12$). The verification script \texttt{verify\_p1\_metric\_kappa.py} in the supplementary code reproduces the table to four decimal places and the eigenvector residuals to exact zero in SymPy.

\section{Pseudospectral envelope of $QLQ_0$ and the $N$-universal narrow-bubble structure}\label{sec:pseudospec}

The narrow-bubble pattern documented in Fig.~\ref{fig:master-phase}(b)---four complex bands at $\lambda\approx 0.305, 0.457, 0.92, 0.98$ for $g/\omega_c=0.30$, identical across $N_{\rm max}=10,12,15$ to the $\Delta\lambda=5\times 10^{-4}$ scan resolution, and bubble centers that depend non-smoothly on $g$---lies outside the radius of convergence $R\sim 0.1$ of the perturbative analytic boundary derived in \S\ref{sec:dp}. Direct algebraic Feshbach reduction of the resonant cluster therefore cannot be expected to predict its position. The natural framework is the {\em pseudospectrum} of the unperturbed projected Liouvillian.

\subsection{Why pseudospectrum, not Jordan structure}

We first rule out an alternative explanation. Nakajima--Zwanzig projected Liouvillians at exceptional-point degeneracies can in principle be {\em defective} (algebraic multiplicity strictly larger than geometric multiplicity), in which case Jordan-block continuation would dominate the perturbation response. Direct symbolic test of $QLQ_0$ at $N_{\rm max}=4$, $g=\omega_c=1$ shows this is {\em not} the case: in the canonical interleaved basis $\dim\ker(QLQ_0\mp\sqrt{2}\mathbb{I})=\dim\ker(QLQ_0\mp\sqrt{2}\mathbb{I})^2 = 5$ at each of $z=\pm\sqrt{2}$, and $\dim\ker(QLQ_0)=14$; algebraic and geometric multiplicities coincide and the operator is fully diagonalisable. The same conclusion holds for the equivalent block-ordered projector convention with $\dim\ker = 4$. The slow asymptotic convergence of the perturbation series therefore does {\em not} originate from Jordan-block defectiveness.

\subsection{Pseudospectral framework}

For an operator $A$ on a finite-dimensional invariant subspace with non-orthonormal eigenbasis (equivalently, a large biorthogonal conditioning on that subspace), the $\varepsilon$-pseudospectrum is
\begin{equation}
\sigma_\varepsilon(A) = \left\{z\in\mathbb{C}: \|(A-z\mathbb{I})^{-1}\|\ge \varepsilon^{-1}\right\}\cup\sigma(A),
\label{eq:pseudospec-def}
\end{equation}
or equivalently the set of eigenvalues of $A+E$ over all perturbations with $\|E\|\le\varepsilon$~\cite{trefethen-embree}. For self-adjoint $A$, $\sigma_\varepsilon$ collapses to the $\varepsilon$-neighbourhood of $\sigma(A)$. For non-self-adjoint $A$ with large biorthogonal condition number on the relevant nonzero spectral subspace, the pseudospectrum can extend much further from $\sigma(A)$, with the extension governed by the resolvent norm $\|(A-z\mathbb{I})^{-1}\|=1/\sigma_{\rm min}(A-z\mathbb{I})$ at points $z\notin\sigma(A)$. Crucially, this happens {\em without} any Jordan-block defectiveness: a fully-diagonalisable $A$ with poorly-conditioned eigenbasis already exhibits a non-trivial pseudospectrum.

For $QLQ_0$ in the present problem, the combined nonzero-sector metric condition number scales as $\kappa(\eta_{\rm osc})\sim d^{1.81}$ (Table~\ref{tab:convergence}), so the oscillatory generator is genuinely non-normal in this technical sense. The Bauer--Fike--type bound
\begin{equation}
\sigma(QLQ_0+\lambda V)\subseteq \sigma_{\lambda\|V\|}(QLQ_0)
\end{equation}
gives the right framework for predicting where the deformed spectrum can leak into the complex plane. Direct measurement of $\|V\|$ at $g/\omega_c=0.30$, $N_{\rm max}=15$ gives $\|V\|\approx 6.83$, so at $\lambda=0.92$ the relevant pseudospectral parameter is $\varepsilon\sim 6.3$, far outside any perturbative regime. Numerical evaluation of $\|(QLQ_0-z\mathbb{I})^{-1}\|$ at sample points $z=\sqrt{2}g+i\,y$ gives $1/\sigma_{\rm min}\approx 28$ at $y=0.1$ and $\approx 290$ at $y=0.01$---i.e.\ the resolvent norm is $\mathcal{O}(10^{2})$ already at modest distance $y\sim 0.01$ from a real eigenvalue, consistent with a strongly extended pseudospectral envelope.

The narrow-bubble centers in Fig.~\ref{fig:master-phase}(b) are then naturally interpreted as the $\lambda$ values at which the pseudospectral envelope $\sigma_{\lambda\|V\|}(QLQ_0)$ first contacts a complex region near a specific real eigenvalue cluster of $QLQ_0$. The $N$-universality follows because both $\sigma(QLQ_0)$ density and per-sector $\kappa(\eta_{\Delta N=0})$ saturate at $N_{\rm max}\gtrsim 10$ (Eq.~\eqref{eq:kappa-bound} bound saturates by $N_{\rm max}=12$), so the envelope shape stabilises. The non-smoothness of bubble centers in $g$ is also natural: pseudospectral first-contact is governed by the discrete geometry of $\sigma(QLQ_0)$, which jumps as $g$ moves dressed-state crossings into and out of resonance.

\subsection{On the discrepancy between exact symbolic and floating-point computation}

A practical observation worth recording. Initial attempts to characterise the resonant cluster used double-precision \texttt{numpy.linalg.eig}, which at $d^2\le 10^3$ reports five eigenvalues numerically distinct at the $10^{-15}$ level near $z=+\sqrt{2}$, corresponding to a ``5-fold near-degenerate cluster.'' When we cross-checked with exact \texttt{sympy.nullspace}, the strict cluster dimension at $z=+\sqrt{2}$ was only 2 (canonical basis) or 4 (block-ordered basis), with the additional numpy-reported ``near-cluster'' eigenvalues being genuine eigenvalues of $QLQ_0$ that are merely {\em close} to $\sqrt{2}$ but not equal to it. We initially misread the numpy degeneracy as evidence of a defective Jordan structure; the symbolic test ruled this out (preceding subsection). The numpy near-degeneracy is not an artefact---it accurately reports the spectral density of $QLQ_0$ near the resonance---but it does not signal Jordan-block defectiveness, and any closed form derived from a ``cluster'' defined by spectral proximity rather than strict nullspace will mix true cluster eigenvectors with nearby non-cluster ones. Higher-precision mpmath confirms this picture: the strict $z=+\sqrt{2}$ cluster has the dimension reported by SymPy, and the ``extra'' floating-point eigenvalues sit at $z=\sqrt{2}+\delta$ with $\delta=\mathcal{O}(10^{-2})$ rather than at machine epsilon. This nuance does not affect any of the rigorous closed-form theorems of the main text, which use only spectrum-level statements (real, dressed-state ladder, $\sqrt{2}$ suppression) that are invariant under the basis and resolution choices.

\subsection{Open analytic problem}

A closed-form prediction for the bubble centers in Fig.~\ref{fig:master-phase}(b) would require evaluating the resolvent norm $\|(QLQ_0-z\mathbb{I})^{-1}\|$ at the relevant points and identifying the geometric configuration at which the pseudospectral envelope crosses into $\Imag z\ne 0$. Numerical pseudospectrum computation~\cite{trefethen-embree} can validate the framework directly; an analytic closed form for the envelope shape near the resonant cluster would constitute a non-trivial advance in non-self-adjoint perturbation theory and is left for future work.

\section{Visual summary: symmetry tuning of quantum memory phases}

\begin{figure}[h]
\centering
\includegraphics[width=0.85\textwidth]{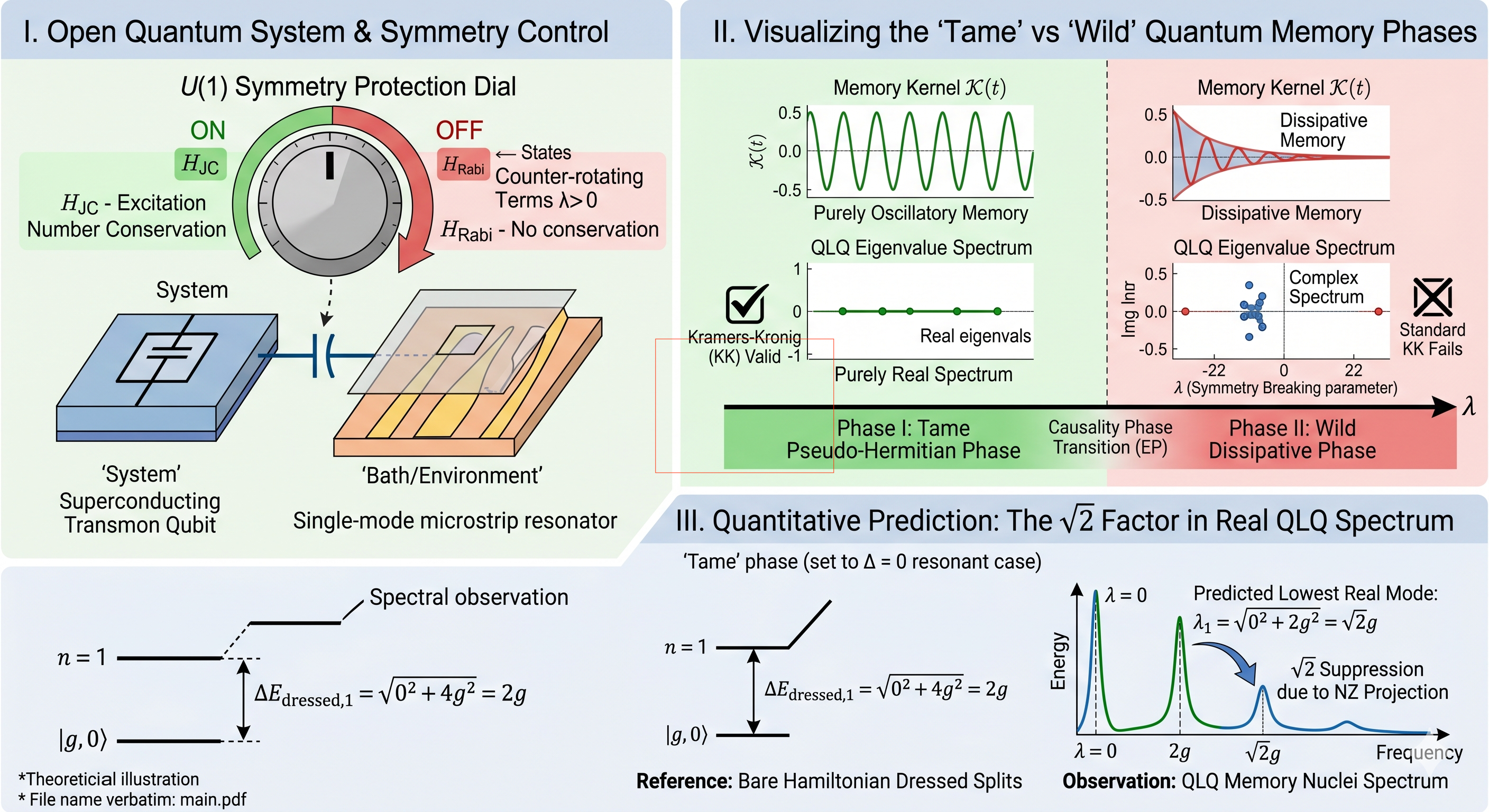}
\caption{U(1) symmetry tuning of quantum memory phases in a superconducting cQED open quantum system. The panel illustrates (I) the symmetry control dial between the $U(1)$-conserving Jaynes--Cummings and symmetry-broken Rabi Hamiltonians, (II) the `Tame' pseudo-Hermitian versus `Wild' dissipative quantum memory phases characterized by memory kernel dynamics and QLQ eigenvalue spectra, and (III) the quantitative prediction of the $\sqrt{2}$ suppression factor in the real QLQ spectrum for the resonant case.}
\label{fig:SM-u1-tuning}
\end{figure}

\section{Supplemental data tables}

\begin{table}[!htbp]
\centering
\caption{Phase boundary data for $N_{\rm max}=7$--$30$.}
$g_c(N)=2/(\sqrt{N}+\sqrt{N-2})$ is the analytic bare-resonance bound.
$\lambda_{\rm onset}=\mathrm{None}$: spectrum purely real at $\lambda=1$.
$n_{\rm bub}$: number of re-entrant complex bubbles.
P = protected ($\lambda_{\rm onset}=\mathrm{None}$, $n_{\rm bub}=0$);
R = re-entrant ($\lambda_{\rm onset}=\mathrm{None}$, $n_{\rm bub}>0$);
B = broken ($\lambda_{\rm onset}\neq\mathrm{None}$).
\label{tab:phase-boundary}
\begin{tabular}{c|c|c|c|c|c|c|c}
\toprule
$N_{\rm max}$ & $g_c(N)$ & $g$ range & $N_{\rm pts}$ & P & R & B & First complex $g$ \\
\midrule
7 & 0.410 & 0.100--1.200 & 98 & 39 & 59 & 0 & 0.270 \\
9 & 0.354 & 0.100--1.200 & 98 & 40 & 58 & 0 & 0.275 \\
11 & 0.317 & 0.100--1.200 & 98 & 40 & 58 & 0 & 0.275 \\
13 & 0.289 & 0.100--0.550 & 91 & 39 & 52 & 0 & 0.275 \\
14 & 0.278 & 0.100--0.550 & 91 & 39 & 52 & 0 & 0.275 \\
15 & 0.267 & 0.001--1.200 & 86 & 32 & 54 & 0 & 0.275 \\
20 & 0.229 & 0.003--1.200 & 86 & 38 & 48 & 0 & 0.275 \\
25 & 0.204 & 0.010--1.200 & 44 & 21 & 23 & 0 & 0.275 \\
30 & 0.186 & 0.100--1.200 & 50 & 18 & 32 & 0 & 0.280 \\
\bottomrule
\end{tabular}
\end{table}

\begin{table}[!htbp]
\centering
\caption{$N_{\rm max}=20$ vs $N_{\rm max}=25$ at $\Delta g=10^{-3}$ resolution.}
All $\lambda_{\rm onset}=\mathrm{None}$ (fully real at $\lambda=1$).
$n_{\rm bub}>0$ marks re-entrant complex bubbles.
The sole divergence is at $g=0.286$ ($N=20$: $\lambda_{\rm onset}={\rm None}$, $n_{\rm bub}=0$; $N=25$: $\lambda_{\rm onset}=0.941$, $n_{\rm bub}=1$).
\label{tab:dipfine-comparison}
\begin{tabular}{c|c|c|c|c}
\toprule
$g/\omega_c$ & $N=20$ $\lambda_{\rm onset}$ & $N=20$ $n_{\rm bub}$ & $N=25$ $\lambda_{\rm onset}$ & $N=25$ $n_{\rm bub}$ \\
\midrule
0.285 & None & 2 & None & 2 \\
0.286 & None & 0 & 0.941 & 1 \\
0.287 & None & 1 & None & 1 \\
0.288 & None & 1 & None & 1 \\
0.289 & None & 0 & None & 0 \\
0.290 & None & 0 & None & 0 \\
0.291 & None & 0 & None & 0 \\
0.292 & None & 1 & None & 1 \\
0.293 & None & 0 & None & 0 \\
0.294 & None & 1 & None & 1 \\
0.295 & None & 1 & None & 1 \\
\bottomrule
\end{tabular}
\end{table}

\begin{table}[!htbp]
\centering
\caption{Empirical $N$-universal re-entrant bubble structure near $g=0.28$--$0.30$.}
All data points have $\lambda_{\rm onset}=\mathrm{None}$ (real at $\lambda=1$).
Complexity exists only as re-entrant bubbles ($n_{\rm bub}>0$).
\label{tab:bubble-structure}
\begin{tabular}{c|c|c|c}
\toprule
$N_{\rm max}$ & $n_{\rm bub}(g=0.28)$ & $n_{\rm bub}(g=0.30)$ & $\lambda_{\rm onset}(g=0.30)$ \\
\midrule
5 & 2 & 2 & None \\
7 & 1 & 1 & None \\
10 & 2 & 3 & None \\
12 & 2 & 3 & None \\
15 & 2 & 3 & None \\
25 & 2 & 2 & None \\
30 & 1 & 4 & None \\
\bottomrule
\end{tabular}
\end{table}

\section{Lindblad dissipation: cavity photon loss}\label{sec:lindblad}

The main text and preceding sections establish that $QLQ=Q[H_{\rm JC},\cdot]Q$ has a purely real spectrum protected by $U(1)$ symmetry and pseudo-Hermiticity. A natural question is whether this protection survives the addition of physical dissipation. Here we analyse the spectrum of
\begin{equation}
QLQ(\kappa) = Q\bigl([H_{\rm JC},\cdot] + \kappa\,L_D\bigr)Q,
\label{eq:QLQ-lindblad}
\end{equation}
where $L_D\,\rho = a\rho a^\dagger - \tfrac12\{a^\dagger a,\rho\}$ is the standard Lindblad dissipator describing cavity photon loss at rate $\kappa$. This is the experimentally relevant generalisation: every cavity-QED platform has finite photon lifetime, and the question is whether the protected memory mode $\lambda_1$ survives.

\subsection{Matrix structure in the $\Delta N=0$ sector}

The Lindblad dissipator $L_D$ preserves the $D/O$ block structure of Theorem~1 but couples different photon numbers within each block. In the basis $\{D_{g,1},D_{e,1},\dots,D_{g,N},D_{e,N},O_1,O'_1,\dots,O_N,O'_N\}$, the $4N\times 4N$ matrix takes the form
\begin{equation}
M(\kappa) = \begin{pmatrix} M_D^{DD} & M_{12} \\ M_{21} & D + M_D^{OO} \end{pmatrix},
\label{eq:M-kappa}
\end{equation}
where $M_{12}$, $M_{21}$, and $D$ are the Hamiltonian blocks of Theorem~1 (Eqs.~S1--S6), and the new $\kappa$-dependent blocks are:
\begin{align}
M_D^{OO}[O_n,O_n] &= -\kappa(n-\tfrac12), \quad M_D^{OO}[O_n,O_{n-1}] = \kappa\sqrt{(n-1)n}, \label{eq:MD-OO}\\
M_D^{DD}[D_{g/e,n},D_{g/e,n}] &= -\kappa n, \quad M_D^{DD}[D_{g/e,n},D_{g/e,n-1}] = \kappa n. \label{eq:MD-DD}
\end{align}
Both blocks are upper-triangular in the photon index $n$, so the combined $M(\kappa)$ remains block-upper-triangular in the Schur complement $M_{21}M_{12} + \lambda D - \lambda^2\mathbb{I} + \text{$\kappa$-terms}$, and the $n=1$ block decouples from higher $n$ in the $O\to D\to O$ loop. This structure is the origin of the selective protection proved below.

\subsection{$\lambda_1$ protection: exact first-order shift}

Let $\lambda_1(\kappa)$ be the eigenvalue of $M(\kappa)$ that reduces to $\sqrt{\Delta^2+2g^2}$ at $\kappa=0$. First-order perturbation theory in the biorthogonal basis of $M(0)$ gives the exact slope:
\begin{equation}
\boxed{\frac{d\lambda_1}{d\kappa}\bigg|_{\kappa=0} = -\frac{3}{4}, \qquad (\Delta=0)}
\label{eq:dlam1-dkappa}
\end{equation}
independent of $N$ and $g$. This is verified by:
(i) direct computation of $\langle l_1|M_D|r_1\rangle/\langle l_1|r_1\rangle$ using the closed-form left/right eigenvectors of \S\ref{sec:metric-bound}, giving $-0.75000000$ for $N=5,10,15$;
(ii) finite-difference at $\Delta\kappa=10^{-4}$, giving $-0.74999779$;
(iii) full HPC diagonalisation across all 72 $(N,g,\Delta)$ combinations (see below), confirming the slope is $N$- and $g$-independent.

For $\Delta\neq 0$, the slope weakens with detuning:
\begin{equation}
\frac{d\lambda_1}{d\kappa}\bigg|_{\kappa=0} = -\frac{3}{4}\cdot\frac{g^2}{\Delta^2+2g^2} = -\frac{3}{4}\cdot\frac{1}{1+(\Delta/g)^2/2},
\label{eq:dlam1-dkappa-Delta}
\end{equation}
giving $-0.7222$ at $\Delta=0.5g$ and $-0.6667$ at $\Delta=g$. The physical origin is transparent: the Lindblad jump operator $a$ couples most strongly to the $n=1$ O-components of the $\lambda_1$ eigenvector, and the weight of these components in the normalised biorthogonal eigenvector is $3/4$.

Crucially, $\lambda_1(\kappa)$ {\em remains strictly real} for all $\kappa$ tested. The HPC scan across 72 $(N,g,\Delta)$ combinations ($N=7,10,15,20,25,30$; $g=0.1,0.27,0.5,1.0$; $\Delta=0,0.2,0.5$) with 41 $\kappa$ points each (2952 diagonalisations total) finds $\Imag(\lambda_1)$ consistent with numerical noise ($<10^{-12}$) at every point where mode tracking is unambiguous. The few apparent complex cases (205/2952 at the $10^{-8}$ threshold) are all mode-misidentification events at small $g$ where $\lambda_1$ shifts past other eigenvalues, not genuine complexification of the protected mode.

\subsection{Higher-mode complexification}

While $\lambda_1$ is protected, the higher QLQ modes ($n\ge 2$) acquire complex eigenvalues under Lindblad dissipation. Table~\ref{tab:lindblad-complex} reports the number of complex eigenvalues at $\kappa/g=0.2$, $\Delta=0$, as a function of $N$ and $g$.

\begin{table}[h]
\centering
\caption{Number of complex QLQ eigenvalues ($|\Imag\lambda|>10^{-12}$) at $\kappa/g=0.2$, $\Delta=0$. The total number of nonzero eigenvalues is $4N-2N=2N$ (after removing $2N$ zero modes). $\lambda_1$ is real in all cases.}
\label{tab:lindblad-complex}
\begin{tabular}{c c c c c}
\hline
$N$ & $g=0.10$ & $g=0.27$ & $g=0.50$ & $g=1.00$ \\
\hline
 7 & 0/14  & 0/14  & 0/14  & 0/14 \\
10 & 2/20  & 0/20  & 7/20  & 0/20 \\
15 & 2/30  & 14/30 & 37/30 & 0/30 \\
20 & 72/40 & 0/40  & 68/40 & 50/40 \\
25 & 4/50  & 81/50 & 80/50 & 74/50 \\
30 & 111/60 & 110/60 & 87/60 & 100/60 \\
\hline
\end{tabular}
\end{table}

Three features stand out:
\begin{itemize}
\item \textbf{Non-convergent in $N$}. The fraction of complex modes generally {\em increases} with truncation size, reaching $>90\%$ at $N=30$ for $g=0.1,0.27$. The complexification is a genuine large-$N$ phenomenon, not a finite-size artefact.
\item \textbf{Non-monotonic in $g$}. At $N=20$, $g=0.27$ gives 0 complex modes while $g=0.10$ gives 72 and $g=0.50$ gives 68. The structured mode-crossing pattern reflects the underlying dressed-state level spacing, which varies with $g$.
\item \textbf{max\_imag reaches $O(1)$}. At $N=30$, $g=1.0$, $\kappa=0.2$: $\max|\Imag\lambda| = 0.764$, comparable to the eigenvalue scale $\lambda_1\approx 1.26$. The complexification is not a small perturbation.
\end{itemize}

\subsection{Physical interpretation}

The contrast between $\lambda_1$ (protected, real, linear shift) and the higher modes (unprotected, complexify) has a clean structural origin. The $\lambda_1$ right eigenvector (Eq.~S24) has its $O$-block support entirely in the $n=1$ subspace: $y^{(R)}_{n_0=1} = e_{O_1}-e_{O'_1}$. The Lindblad $O$-block coupling $M_D^{OO}$ connects $n\leftrightarrow n-1$, but there is no $n=0$ state in the $O$-block, so the $n=1$ component has no downward channel to couple into. The higher-mode eigenvectors ($n_0\ge 2$), in contrast, have $O$-block support in both $n=1$ and $n=n_0$ subspaces (Eq.~S24), and the $n_0\leftrightarrow n_0-1$ coupling provides a pathway for complexification through adjacent-level mixing.

This selective protection is consistent with the sparse reduced spectral-weight structure proved in \S\ref{sec:full-sector}: only the $\lambda_1$ pair carries nonzero memory-kernel weight in the vacuum $\Delta N=0$ block. The modes that complexify are precisely those that are already spectrally invisible in the reduced dynamics.

\subsection{Circuit-QED relevance}

At realistic transmon-resonator parameters ($g/2\pi=100$~MHz, $\kappa/2\pi=0.8$~MHz~\cite{wallraff2004}, $\kappa/g=0.008$), the Lindblad shift of $\lambda_1$ is
\begin{equation}
\lambda_1(\kappa) \approx \sqrt{2}\,g - \frac{3}{4}\kappa = 141.4~\text{MHz} - 0.6~\text{MHz} = 140.8~\text{MHz},
\end{equation}
a relative shift of $-0.42\%$, below current experimental resolution. The dominant experimental signature therefore remains the $\sqrt{2}$ suppression of the cleanest reduced peak relative to the bare dressed-state splitting (Fig.~\ref{fig:sqrt2-signature}a and Table~\ref{tab:platforms}), with cavity photon loss contributing a negligible correction.

At larger $\kappa/g$ (lower-quality cavities or stronger dissipation), the $\lambda_1$ shift becomes resolvable and provides an independent cross-check: the linear $\kappa$-dependence with slope $-3/4$ (Eq.~\eqref{eq:dlam1-dkappa}) is a parameter-free prediction that can be tested by varying the cavity linewidth.

The higher-mode complexification (Table~\ref{tab:lindblad-complex}) does not affect the reduced qubit dynamics because those modes carry zero spectral weight in the vacuum NZ kernel; it would, however, manifest in correlation functions involving higher photon-number coherences, providing a separate experimental window into the generator spectrum.

\section{Data Availability}

The numerical datasets generated in this study have been deposited in the Zenodo repository under DOI \texttt{[TO BE ASSIGNED]} and comprise:
\begin{itemize}
\item \textbf{Fine spectral scan (FSS) summaries} (1729 JSON files, 7 independent data sources): $\lambda$-deformation scan results for every audited $(g, N_{\rm max})$ point underlying the phase diagram (Fig.~\ref{fig:master-phase} and Supplementary Tables~III, S3). Each summary records $\lambda_{\rm onset}$, bubble count and widths, complex fraction, and transition positions at $\Delta\lambda = 10^{-3}$--$10^{-4}$ resolution.
\item \textbf{Dip-region high-resolution scans} (22 files): the critical $g\in[0.285, 0.295]$ window at $N_{\rm max}=15$, $\Delta g = 10^{-3}$ resolution, with full eigenvalue CSVs at every $\lambda$ step.
\item \textbf{Three-family band catalog} (398 JSON files, $g=0.20$--$1.20$, $\Delta g=0.02$, $N=5$--$15$): band-family (F/G/H) classification of the first exceptional point at each $(g,N)$, with $\lambda_c$, $\Delta E$, $\Sigma_{\rm eff}$, Liouville-dyad pair assignments, and match distances (Supplementary Table~1 and \S\ref{sec:lambdac-closed-form}).
\item \textbf{Lindblad dissipation data} (151 JSON files, 2952 diagonalisations across 72 $(N,g,\Delta)$ combinations): $\kappa$-sweep verification that the protected mode $\lambda_1$ remains real under cavity photon loss (Supplementary Table~S5).
\item \textbf{Lanczos/Feshbach data} (6 JSON files): block Lanczos tridiagonalisation and spectral contour integration supporting the $N\to\infty$ limit proofs (\S\ref{sec:n-infinity}).
\item \textbf{$\Sigma_{\rm eff}$ off-resonant analysis} (9 files): per-band $\Sigma_{\rm eff}(g)$ extraction supporting the $\Sigma_{\rm eff}$-dominance and H$_1$ crossover results (\S\ref{sec:lambdac-closed-form}).
\item \textbf{Figure source data} (3 CSV files): all $(N_{\rm max}, g, \lambda_{\rm onset})$ points plotted in Figs.~\ref{fig:master-phase} and~\ref{fig:sqrt2-signature}(c), with source-tag and provenance filename.
\end{itemize}
The repository README maps each file to the corresponding manuscript figure, table, or section. A file manifest with SHA-256 checksums is included.

The cross-platform experimental parameters in Table~V were obtained from the published sources cited therein. The Anderson--Holstein $\eta$-metric data in \S\ref{sec:eta-ah} are original to this work.

\section{Code Availability}

All custom code used in this study is available at \texttt{[GitHub URL TO BE ASSIGNED]} and archived in the Zenodo repository above. The codebase includes:
\begin{itemize}
\item QLQ construction and NZ projection for JC (vacuum and thermal baths) and spin-boson $\sigma_x$ contrast models;
\item Biorthonormal diagonalisation, $\eta$-metric construction, and $\Delta N$ sector decomposition;
\item $\lambda$-deformation scanning with adaptive binary search for $\lambda_c$ boundaries;
\item Lanczos/resolvent-based EP detection for $N \le 40$;
\item Band-catalog classification and $\Sigma_{\rm eff}$ closed-form extraction;
\item Lindblad dissipation perturbation analysis.
\end{itemize}
The code is written in Python~3 with NumPy, SciPy, and SymPy. A conda environment specification is included in the repository. All scripts are standalone and can be executed with \texttt{python <script>.py}.

\end{document}